\documentclass[superscriptaddress,twocolumn,aps,pra,preprintnumbers,notitlepage,showpacs,nofootinbib]{revtex4-1}
\usepackage{amsmath,amssymb}
\usepackage{lipsum} 
\usepackage[latin9]{inputenc}
\usepackage{graphicx}
\usepackage{dcolumn}
\usepackage{bbold}
\usepackage{bm}
\usepackage[usenames]{xcolor}
\usepackage[colorlinks,bookmarks=false,citecolor=blue,linkcolor=blue,urlcolor=blue,hyperfootnotes=true]{hyperref}
\usepackage[caption=false,labelformat=simple]{subfig}
\usepackage{accents}

\begin{document}
\title{Resistivity near a nematic quantum critical point: Impact of acoustic phonons}
\author{V. S. de Carvalho}
\email{vsilva@ifi.unicamp.br}
\affiliation{School of Physics and Astronomy, University of Minnesota, Minneapolis,
Minnesota 55455, USA}
\affiliation{Instituto de F\'{i}sica Gleb Wataghin, Unicamp, 13083-859, Campinas-SP,
Brazil}
\author{R. M. Fernandes}
\affiliation{School of Physics and Astronomy, University of Minnesota, Minneapolis,
Minnesota 55455, USA}

\date{\today}
\begin{abstract}
We revisit the issue of the resistivity of a two-dimensional
electronic system tuned to a nematic quantum critical point (QCP), focusing on the non-trivial impact of the coupling to the acoustic phonons. Due to the unavoidable linear coupling between
the electronic nematic order parameter and the lattice strain fields, long-range nematic interactions mediated by the phonons emerge in the problem. By solving the semi-classical Boltzmann equation in the presence of scattering by impurities and nematic fluctuations, we determine the temperature dependence of the resistivity as the nematic QCP is approached. One of the main effects of the nemato-elastic coupling is to smooth the electronic non-equilibrium distribution function, making it approach the simple cosine angular dependence even when the impurity scattering is not too strong. We find that at temperatures lower than a temperature scale set by the nemato-elastic coupling, the resistivity shows the $T^2$ behavior characteristic of a Fermi liquid. This is in contrast to the $T^{4/3}$ low-temperature behavior expected for a lattice-free nematic quantum critical point. More importantly, we show that the effective resistivity exponent $\alpha_\text{eff}(T)$ in $\rho(T)-\rho_0\sim T^{\alpha_\text{eff}(T)}$ displays a pronounced temperature dependence, implying that a nematic QCP cannot generally be characterized by a simple resistivity exponent. We discuss the implications of our results to the interpretation of experimental data, particularly in the nematic superconductor FeSe$_{1-x}$S$_x$.
\end{abstract}

\maketitle

\section{Introduction}

\label{Sec_I}

Recent experiments in several quantum materials have revealed the
widespread presence of electronic nematicity, i.e. the lowering of
the crystalline point-group symmetry by electronic degrees of freedom \cite{Hinkov-S(2008),Taillefer-N(2010),Hashimoto-S(2012),Fujita-S(2014),Hosoi-PNAS(2016),Kuo-S(2016),Matsuda-NP(2017),Zheng-PRL(2018),Coldea-ARCMP(2018),Bohmer-JPCM(2018),Ronning17}.
Assessing the impact of these nematic degrees of freedom on the normal-state
and superconducting properties of these materials remains an important
challenge, particularly near a putative nematic quantum critical point
(QCP) \cite{Fradkin-ARCMP(2010),Metlitski-PRB(2010),Senthil-PRB(2010),Kopietz-PRB(2012),Metzner-PRB(2015),Metlitski-PRB(2015),Berg-PRX(2016),Lederer-PNAS(2017),Lee-ARCMP(2018)}. Experiments provide some evidence that the nematic transition
can indeed be tuned to zero temperature by doping or pressure \cite{Coldea-ARCMP(2018),Bohmer-JPCM(2018),Licciardello-N(2019)}. However, progress
in elucidating the properties of a nematic quantum critical metal
is often hindered by the fact that other types of ordered states are
observed concomitantly, such as magnetic order in the pnictides \cite{Fernandes-NP(2014)}, or charge order in the cuprates \cite{Keimer-N(2015)}. The simultaneous
presence of fluctuations associated with multiple ordered phases makes
it difficult to disentangle the relevance of the putative nematic
QCP to the non-Fermi liquid behavior or to the unconventional superconducting
dome often observed in these systems.

However, materials have been recently discovered that seem to display
only nematic order, disentangled from other ordered states. This is
the case of the chemically substituted iron chalcogenide FeSe$_{1-x}$S$_{x}$ \cite{Hosoi-PNAS(2016)}:
for $x=0$, the system displays a nematic transition at $T_{\mathrm{nem}}\approx90$
K, whereas for $x=1$ the system is tetragonal. A putative nematic
QCP is inferred near the $x\approx0.18$ concentration, although it
is not clear whether the transition is first-order (in which case
the QCP would be avoided) or second-order. A nickel-based cousin of
the iron pnictides, Ba$_{1-x}$Sr$_{x}$Ni$_{2}$As$_{2}$, and LaFeAsO$_{1-x}$F$_x$ also show
evidence for a putative nematic QCP without magnetic order \cite{Eckberg-arXiv(2019),Yang-CPL(2015)}, although
for the former charge fluctuations may be important. Finally, certain $4f$ intermetallics,
such as TmAg$_{2}$, undergo a single transition to a nematic phase as temperature
is lowered \cite{Morin-PRB(1993)}. It has been proposed that shear strain
can be used to tune this nematic transition to zero temperature \cite{Maharaj-PNAS(2017)}, promoting
a putative nematic QCP.

These observations motivate a closer theoretical investigation of
the metallic nematic QCP and, particularly, of its transport properties,
since resistivity is one of the most widely employed probes for non-Fermi
liquid behavior. Within the so-called Hertz-Millis approach \cite{Hertz-PRB(1976),Millis-PRB(1993),Lohneysen-RMP(2007)}, in which
electronic degrees of freedom are integrated out, the dynamic exponent
$z$ characterizing the nematic QCP is the same as that of a ferromagnetic
QCP, $z=3$ \cite{Pepin06}. This is because Landau damping has the same form in both
cases, since the two ordered states have zero wave-vector. A semi-classical
Boltzmann-equation approach then predicts that, for a two-dimensional
system, the resistivity $\Delta\rho(T)\equiv\rho(T)-\rho_{0}$, where $\rho_{0}$
is the residual resistivity, vanishes as the QCP is approached according
to $\Delta\rho(T)\sim T^{4/3}$ \cite{Metzner-PRL(2007),*Metzner-PRL(2009),Lohneysen-RMP(2007)}.
Importantly, the presence of impurity scattering to provide a mechanism
for momentum relaxation and multiple bands to avoid geometrical cancellation
effects are essential \cite{Maslov-PRL(2011),Pal-LJPTS(2012)}. Such
an exponent, however, has not been observed in recent transport measurements
in ``optimally doped'' FeSe$_{1-x}$S$_{x}$ \cite{Licciardello-N(2019)} \textendash{} in contrast,
the resistivity of certain metallic ferromagnets near the QCP seems
to be consistent with Hertz-Millis predictions \cite{Lohneysen-RMP(2007)}. There are several
reasons that could be behind this disagreement, from the possible
unsuitability of the Boltzmann-equation approach to describe a system
without well-defined quasi-particles to the possible failure of the
Hertz-Millis description. Indeed, calculations using the memory matrix
formalism have found different temperature dependencies for $\Delta\rho(T)$ \cite{Hartnoll-PRB(2014),Wang-PRB(2019)}.
Quantum Monte Carlo simulations also provide evidence for non-Hertz-Millis
behavior near a nematic QCP \cite{Berg-PRX(2016),Lederer-PNAS(2017)}.

Although both the ferromagnetic and nematic QCPs are characterized
by the same dynamic exponent $z=3$ in the Hertz-Millis approach,
a crucial distinction between them is that the nematic order parameter
couples linearly to elastic modes of the tetragonal lattice \cite{Xu_Qi2009,Fernandes2010,Paul-PRB(2010),Dagotto2013,Schmalian-PRB(2016),Paul-PRL(2017),Paul-PRB(2017)}, which
are associated with acoustic phonon modes. As a result, the acoustic phonons mediate long-range interactions involving the nematic
order parameter \cite{Cowley-PRB(1976)}. While these interactions render the classical (i.e.,
thermal) nematic transition mean-field like, they also restore Fermi-liquid
like thermodynamic behavior near the nematic QCP, as shown in Ref. \cite{Paul-PRL(2017)}.

In this paper, we focus on the impact of the coupling to elastic degrees
of freedom on the transport properties of  a two-dimensional (2D) electronic system close to a nematic
QCP. Because this coupling promotes well-defined quasi-particles near
the QCP, we employ a Hertz-Millis Boltzmann-equation approach to
calculate the temperature-dependence of the resistivity $\Delta\rho(T)$
upon approaching the QCP \cite{Ziman-OUP(1960),Hlubina-PRB(1995),Rosch-PRL(1999),Fernandes2012}. We go beyond the relaxation-time approximation
by solving numerically the Boltzmann equation, from which we obtain
the non-equilibrium electronic distribution function. Impurity scattering
is included as the main source for electronic momentum relaxation.
We obtain a momentum-anisotropic distribution function due to the
interplay between the $d$-wave nematic form factor and the coupling
to the elastic degrees of freedom. The main effect of the latter is
to cause the nematic correlation length to diverge only along certain
momentum-space directions, as discussed previously in Refs. \cite{Schmalian-PRB(2016),Paul-PRL(2017)}. At the lowest
temperatures, we obtain the standard Fermi-liquid-like behavior $\Delta\rho(T)\sim T^{2}$,
which is consistent with the Fermi-liquid behavior previously found
in equilibrium thermodynamic properties  of the same model \cite{Paul-PRL(2017)}. But our key result is that,
upon approaching the nematic QCP, the resistivity cannot be described
by a simple power-law $\Delta\rho(T)\sim T^{\alpha}$ over a wide temperature
range. Instead, the effective temperature-dependent exponent $\alpha_{\mathrm{eff}}(T)\equiv\partial\ln\left[\Delta\rho(T)\right]/\partial\ln T$
displays a prominent temperature dependence, crossing over from $4/3$
at moderate temperatures to $2$ at very low temperatures. This regime
in which $\alpha_{\mathrm{eff}}(T)$ is strongly temperature dependent
is particularly sizable for systems with strong nemato-elastic coupling
and small Fermi energy, as it is presumably the case of the iron-based
superconductors. We also contrast our theoretical results with recent
experiments performed in FeSe$_{1-x}$S$_{x}$ \cite{Coldea-arXiv(2019),Licciardello-N(2019)}, and discuss the limitations
of our approach.

This paper is structured as follows. In Sec. \ref{Sec_II}, we describe
the 2D electronic model for the nematic QCP coupled to elastic degrees
of freedom. In Sec. \ref{Sec_III}, we give a brief description of
the formalism involved in the derivation of the Boltzmann equation
and present its numerical solution as a function of temperature
as well as other parameters of the model. After that, we describe
the low-temperature behavior of the resistivity obtained
from the solution of the Boltzmann equation. Lastly, Sec. \ref{Sec_IV}
is devoted to the discussion of the results and the presentation of
our concluding remarks. The details of some numerical calculations are given in the Appendix.

\section{Microscopic model}

\label{Sec_II}

We consider here a two-dimensional tetragonal electronic
system coupled to nematic quantum fluctuations and the
elastic degrees of freedom of the lattice, similar to that in
to Ref. \cite{Paul-PRL(2017)}. The Hamiltonian
is given by $\mathcal{H}=\mathcal{H}_{\text{el-nem}}+\mathcal{H}_{\text{nem-latt}}$,
where $\mathcal{H}_{\text{el-nem}}$ describes the coupling between
the fermions and the $B_{1g}$ nematic order parameter $\phi(\mathbf{q})$,
whereas $\mathcal{H}_{\text{nem-latt}}$ denotes, from a renormalization-group point of view, the most relevant coupling of $\phi(\mathbf{q})$
to the local orthorhombic strain $\epsilon(\mathbf{r})=\epsilon_{xx}(\mathbf{r})-\epsilon_{yy}(\mathbf{r})$.
Here, strain is defined in the standard way in terms of the displacement
vector $\bm{u}$, such that $\epsilon_{ij}=\partial_{i}u_{j}+\partial_{j}u_{i}$ \cite{Landau-PP(1970),Cowley-PRB(1976)}.
To keep the analysis as simple as possible, we consider a single circular Fermi surface that has nematic cold spots, with dispersion $\xi_{\mathbf{k}}=\varepsilon_{\mathbf{k}}-\mu$,
where $\mu$ is the chemical potential, and write $\mathcal{H}_{\text{el-nem}}$
according to 
\begin{widetext}
\begin{equation}
\mathcal{H}_{\text{el-nem}}=\sum\limits _{\mathbf{k},\sigma}\xi_{\mathbf{k}}\psi_{\sigma}^{\dagger}(\mathbf{k})\psi_{\sigma}(\mathbf{k})+\frac{g_{\text{nem}}}{\sqrt{\nu_{0}}}\sum\limits _{\mathbf{k},\mathbf{q},\sigma}h_{\mathbf{k}}\psi_{\sigma}^{\dagger}(\mathbf{k}+\mathbf{q}/2)\psi_{\sigma}(\mathbf{k}-\mathbf{q}/2)\phi(\mathbf{q}),\label{Eq_Nem_Elec}
\end{equation}
where $\psi_{\sigma}^{\dagger}(\mathbf{k})$ {[}$\psi_{\sigma}(\mathbf{k})${]}
creates (annihilates) electrons with momentum $\mathbf{k}$ and spin
   projection $\sigma\in\{\uparrow,\downarrow\}$, $g_{\text{nem}}$
is the nematic coupling, and $h_{\mathbf{k}}$ denotes the $d$-wave
nematic form factor. We introduced the density of states $\nu_{0}$
for convenience. In the case of a $B_{1g}$ nematic instability,
in which the tetragonal symmetry is broken by making the $x$ and
$y$ directions inequivalent, $h_{\mathbf{k}}$ is given by:
\begin{equation}
h_{\mathbf{k}}=\cos(k_{x})-\cos(k_{y}),
\end{equation}
where the momentum $\mathbf{k}$ is restricted to the vicinity of the
Fermi surface. Note that $h_{\mathbf{k}}$ vanishes along the diagonals of the Brillouin zone.
Thus, the electronic states at the points where the Fermi surface
intercepts these diagonals are effectively uncoupled from the nematic
fluctuations. For this reason, they are known as \emph{cold spots}. 

The nematic degrees of freedom are described by the bare propagator:
\begin{equation}
\left(\chi_{\mathrm{nem}}^{0}\right)^{-1}\left(\mathbf{q},i\Omega_{n}\right)=\nu^{-1}_0\bigg(r_{0}+q^{2}+\frac{\Omega_{n}^{2}}{c^{2}}\bigg),
\end{equation}
where $c$ is a constant, $\Omega_{n}=2n\pi T$ for $n\in\mathbb{Z}$ is the bosonic Matsubara
frequency, and $r_{0}$ is the control parameter proportional to the
distance to the bare nematic QCP. Hereafter, all momenta are given
in units of the inverse lattice constant, whereas all length scales
are given in units of the lattice constant.

As for the Hamiltonian $\mathcal{H}_{\text{nem-latt}}$, it is given
by 
\begin{equation}
\mathcal{H}_{\text{nem-latt}}=\frac{1}{2}\sum\limits _{\mathbf{q}\neq0}^ {}\bm{u}^{\dagger}(\mathbf{q})\mathcal{\boldsymbol{M}}(\mathbf{q})\bm{u}(\mathbf{q})+ig_{\mathrm{latt}}\sum\limits _{\mathbf{q}\neq0}\mathbf{a}_{\mathbf{q}}\cdot\bm{u}(\mathbf{q})\phi(-\mathbf{q}),\label{H_nem_latt}
\end{equation}
where $\bm{u}(\mathbf{q})$ denotes the Fourier transform of the displacement
vector, $\mathbf{a}_{\mathbf{q}}=(q_{x},-q_{y},0)$ is a two-dimensional
vector, $g_{\mathrm{latt}}$ represents the nemato-elastic coupling,
and $\mathcal{\boldsymbol{M}}(\mathbf{q})$ stands for the matrix:  
\begin{equation}
\mathcal{\boldsymbol{M}}(\mathbf{q})=\begin{pmatrix}C_{11}q_{x}^{2}+C_{66}q_{y}^{2}+C_{44}q_{z}^{2} & (C_{12}+C_{66})q_{x}q_{y} & (C_{13}+C_{44})q_{x}q_{z}\\
(C_{12}+C_{66})q_{x}q_{y} & C_{66}q_{x}^{2}+C_{11}q_{y}^{2}+C_{44}q_{z}^{2} & (C_{13}+C_{44})q_{y}q_{z}\\
(C_{13}+C_{44})q_{x}q_{z} & (C_{13}+C_{44})q_{y}q_{z} & C_{44}(q_{x}^{2}+q_{y}^{2})+C_{33}q_{z}^{2}
\end{pmatrix},
\end{equation}
with the $C_{ij}$ denoting the elastic constants for a system with
tetragonal symmetry \cite{Landau-PP(1970),Cowley-PRB(1976)}.
\end{widetext}

The effect of the elastic coupling on the nematic degrees of freedom
can be evaluated by integrating out the $\bm{u}(\mathbf{q})$ fields.
Following the procedure outlined in Ref. \cite{Paul-PRL(2017)}, we first project $\bm{u}(\mathbf{q})$
onto the basis of the polarization vectors of the acoustic phonons
$\hat{\bm{e}}_{\mu}(\mathbf{q})$:
\begin{equation}
\bm{u}(\mathbf{q})=\sum_{\mu}U_{\mu}(\mathbf{q})\hat{\bm{e}}_{\mu}(\mathbf{q}).
\end{equation}
The polarization vectors are given by the eigenvalue equation $\mathcal{\boldsymbol{M}}(\mathbf{q})\hat{\bm{e}}_{\mu}(\mathbf{q})=\varrho\omega_{\mu}^{2}(\mathbf{q})\hat{\bm{e}}_{\mu}(\mathbf{q})$,
where $\omega_{\mu}(\mathbf{q})$ are the phonon dispersions and $\varrho$
is the mass density. Next, we use a path integral representation for
$\mathcal{H}_{\text{nem-latt}}$ and then integrate out the elastic
degrees of freedom. As a result, we obtain the renormalized nematic
propagator
\begin{equation}
\chi_{\mathrm{nem}}^{-1}\left(\mathbf{q},i\Omega_{n}\right)=\left(\chi_{\mathrm{nem}}^{0}\right)^{-1}\left(\mathbf{q},i\Omega_{n}\right)-\Pi_{\mathrm{latt}}(\mathbf{q},i\Omega_{n}),\label{nem_suscept}
\end{equation}
with the polarization bubble: 
\begin{equation}
\Pi_{\text{latt}}(\mathbf{q},i\Omega_{n})=\frac{g_{\mathrm{latt}}^{2}}{\varrho}\sum_{\mu}\frac{|\mathbf{a}_{\mathbf{q}}\cdot\hat{\bm{e}}_{\mu}(\mathbf{q})|^{2}}{\omega_{\mu}^{2}(\mathbf{q})+\Omega_{n}^{2}}.\label{Eq_Latt_Bubble}
\end{equation}

This last expression, when Fourier-transformed back to real space, corresponds to a long-range interaction experienced by 
the nematic degrees of freedom $\phi(\mathbf{q})$ \cite{Paul-PRL(2017)}.
To proceed, since we are mostly interested in layered systems, we
hereafter set $q_{z}=0$. By defining the angle $\varphi=\tan^{-1}(q_{y}/q_{x})$
between the two components of $\mathbf{q}$ restricted to the $x$-$y$
plane and the elastic constants combinations $\gamma_{1}=C_{11}+C_{66}$,
$\gamma_{2}=C_{11}-C_{66}$, and $\gamma_{3}=C_{12}+C_{66}$, we find
that the eigenvalues of $\mathcal{\boldsymbol{M}}(\mathbf{q})$ can
be written as $\omega_{\mu}(\mathbf{q})=v_{\mu}(\varphi)|\mathbf{q}|$,
with anisotropic sound velocities \cite{Schutt-PRB(2018)}:
\begin{align}
v_{\pm}(\varphi) & =\frac{1}{\sqrt{2\varrho}}\sqrt{\gamma_{1}\pm\sqrt{\frac{\gamma_{2}^{2}+\gamma_{3}^{2}}{2}+\frac{\gamma_{2}^{2}-\gamma_{3}^{2}}{2}\cos(4\varphi)}},\label{Eq_EValues_A}
\end{align}
Accordingly, the normalized eigenvectors of that matrix are given by: 
\begin{align}
\hat{\bm{e}}_{\pm}(\varphi) & =\frac{1}{\sqrt{2\Upsilon[\Upsilon\mp\gamma_{2}\cos(2\varphi)]}}\begin{pmatrix}\gamma_{3}\sin(2\varphi)\\
\pm\Upsilon-\gamma_{2}\cos(2\varphi)\\
0
\end{pmatrix},\label{Eq_EVectors_A}
\end{align}
where $\Upsilon\left(\varphi\right)=\sqrt{\frac{1}{2}(\gamma_{2}^{2}+\gamma_{3}^{2})+\frac{1}{2}(\gamma_{2}^{2}-\gamma_{3}^{2})\cos(4\varphi)}$
is a $C_{4}$-symmetric function.

\begin{figure}[t]
\centering
\includegraphics[width=0.95\linewidth]{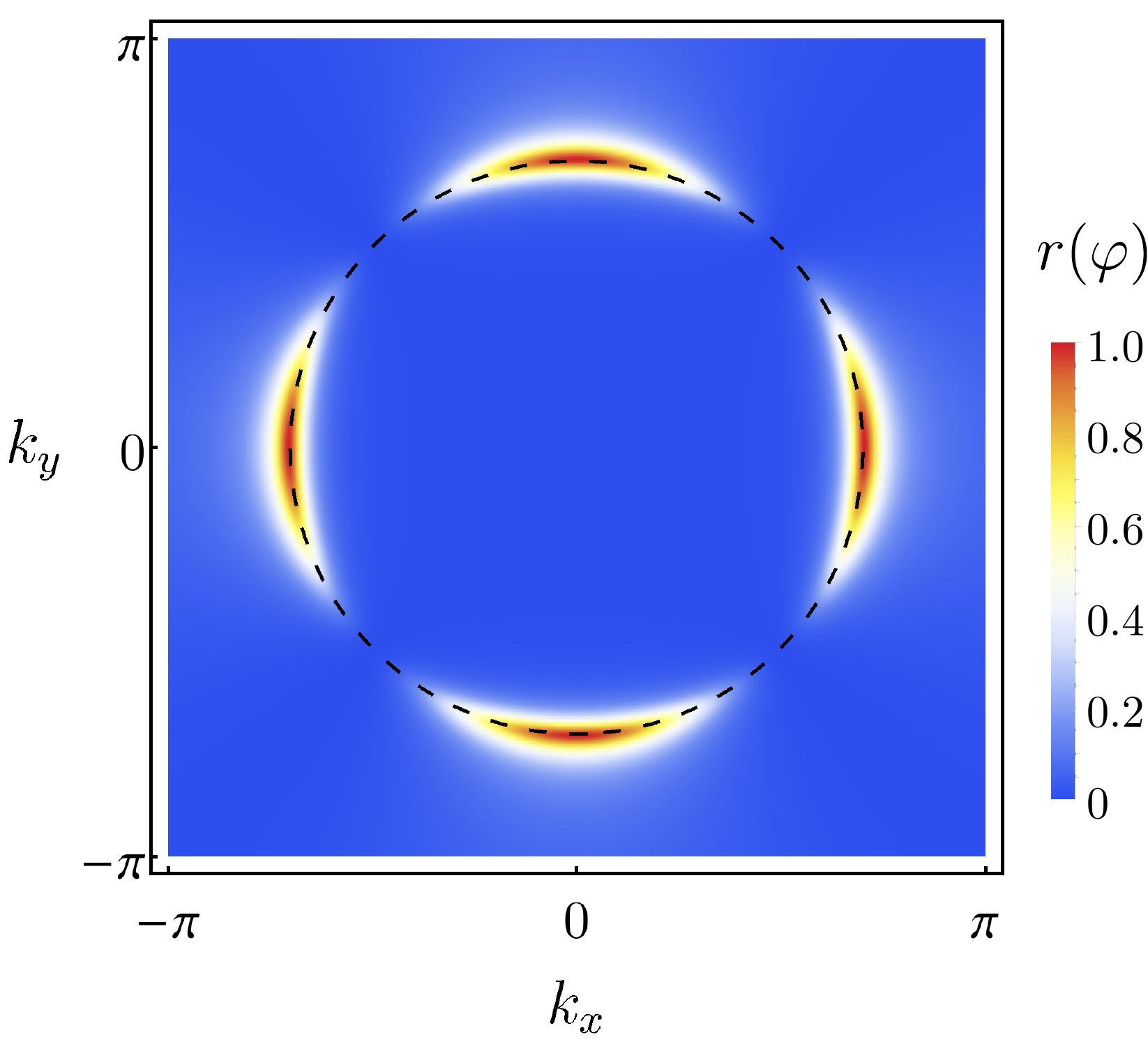}
\caption{(Color online) Density plot of the effective mass $r(\varphi)$ (in units of $\lambda_\text{latt}$) of the critical nematic fluctuations around a circular Fermi surface (dashed black circumference). Notice that  $r(\varphi)$ only goes to zero at the cold spots located at the intersection of the Brillouin-zone diagonals with the Fermi surface. In this situation, the quantum critical regime becomes directional-selective.}\label{Fermi_Surface} 
\end{figure}

By inserting Eqs. \eqref{Eq_EValues_A}\textendash \eqref{Eq_EVectors_A}
into Eq. \eqref{Eq_Latt_Bubble}, we can determine straightforwardly
the static polarization bubble $\Pi_{\text{latt}}(\mathbf{q},i\Omega_{n}=0)$
and, consequently, the effective mass of the nematic fluctuations
defined according to $\nu^{-1}_0r\equiv\chi_{\mathrm{nem}}^{-1}(0,0)$. We obtain
\begin{equation}
r(\varphi)=r_{0}-\frac{2g_{\mathrm{latt}}^{2}\nu_0}{\gamma_{1}^{2}-\Upsilon^{2}\left(\varphi\right)}\bigg[\gamma_{1}+\gamma_{3}-\left(\gamma_{2}+\gamma_{3}\right)\cos^{2}(2\varphi)\bigg].\label{Eq_Ren_Mass}
\end{equation}
Note that $r(\varphi)$ is invariant under $\pi/2$ rotations, and
has a set of minima for $\varphi_{n}=(2n+1)\pi/4$ with $n\in\mathbb{Z}$.
It is convenient to consider the case in which $\frac{C_{11}-C_{12}}{2}=C_{66}$,
which corresponds to a lattice that is equally hard with respect to
the two types of orthorhombic distortion fluctuations. In this case, one has
$\gamma_{2}=\gamma_{3}$ and, consequently, the expression above becomes: 
\begin{equation}
r(\varphi)=r_{0}-r_{0,c}+\lambda_\text{latt}\cos^{2}(2\varphi),\label{Eq_Effective_Mass}
\end{equation}
with $r_{0,c}\equiv\frac{g_{\mathrm{latt}}^{2}\nu_0}{C_{66}}$ and $\lambda_\text{latt}\equiv\frac{r_{0,c}}{2}\left(1+\frac{C_{12}}{C_{11}}\right)$
being two positive parameters. Thus, the elastic degrees of freedom
have two effects: Firstly, they shift the nematic QCP from $r_{0}=0$ to $r_{0}=r_{0,c}>0$
and, secondly, they endow the effective nematic mass with an angular dependence.
As a result, the nematic correlation length $\xi_{\mathrm{nem}}\propto r^{-1/2}$
only diverges along the special momentum-space directions $\varphi_{n}=(2n+1)\pi/4$,
which coincide with the cold spots positions. This is illustrated
in Fig. \ref{Fermi_Surface}. For this reason, the quantum critical
regime becomes directional selective, as discussed in Ref. \cite{Paul-PRL(2017)}. This
non-analytic behavior in momentum space can be recast in terms of
long-range, dipolar-like interactions in real space involving the
nematic order parameter \cite{Xu_Qi2009,Schmalian-PRB(2016)}.

\section{Resistivity near the nematic QCP}

\label{Sec_III}

The consequences of the directional-dependent nematic correlation
length, Eq. \eqref{Eq_Ren_Mass}, to the thermodynamic and single-particle
electronic properties near the nematic QCP were established in Ref.
\cite{Paul-PRL(2017)}. The main result is that, at low enough temperatures, electronic
quasi-particles are well-defined, and the system displays a conventional
Fermi-liquid behavior. Of course, this temperature scale depends crucially
on the nemato-elastic coupling constant $g_{\mathrm{latt}}$. Our
goal here is to determine the transport properties near the QCP. While
one may be tempted to employ a relaxation-time approximation and replace
the transport lifetime by the single-particle lifetime, it is well
known that this approximation can be problematic in the case of anisotropic
scattering \cite{Hlubina-PRB(1995),Rosch-PRL(1999),Rosch-PRB(2000),Fernandes2012}. Furthermore, the relaxation-time approximation makes no reference to momentum relaxation mechanisms, which are particularly
important near instabilities with zero wave-vector \cite{Maslov-PRL(2011)}.

\subsection{Boltzmann equation formalism}

\begin{figure}[t]
\centering
\includegraphics[width=0.90\linewidth]{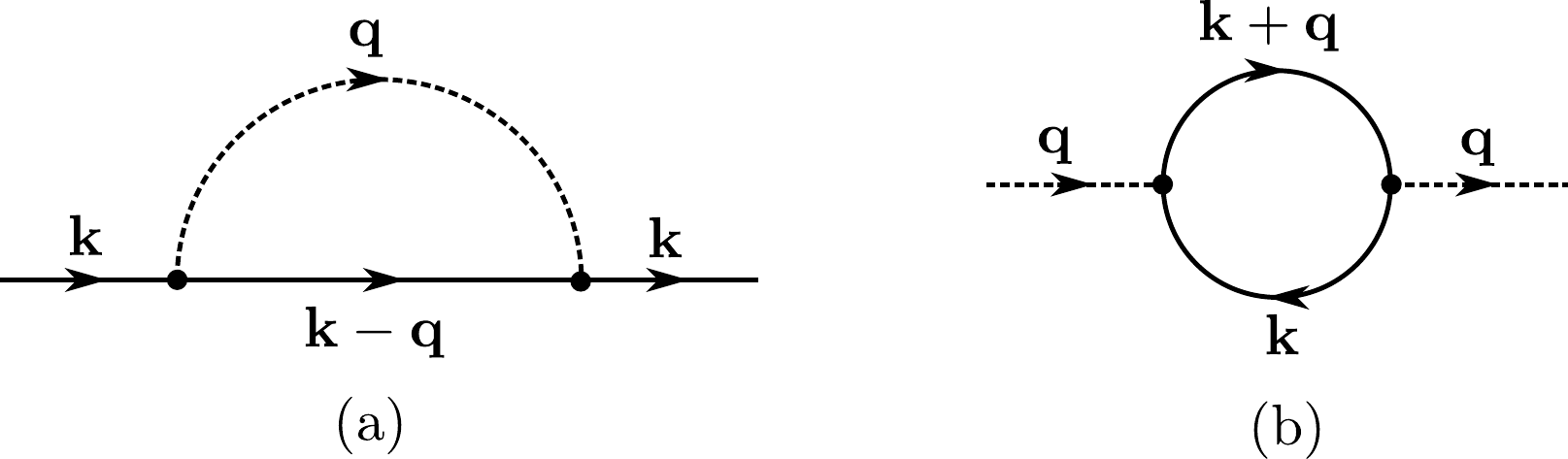}
\caption{Self-energy corrections for (a) the electronic and (b) the nematic
degrees of freedom. The electronic and bosonic propagators are denoted
here by the solid and dashed lines, respectively. The black dots
(``$\bullet$") correspond to the product of the nematic
interaction $g_\text{nem}$ with the nematic $d$-wave form factor $h_\mathbf{k}$.}\label{One_Loop} 
\end{figure}

We calculate the electrical resistivity by employing a semi-classical
Boltzmann-equation approach \cite{Ziman-OUP(1960),Hlubina-PRB(1995),Rosch-PRL(1999),Fernandes2012}.
Because the coupling to the elastic degrees of freedom restores well-defined
quasi-particles at the QCP, as discussed above, such a semi-classical
approach is not unreasonable. We will come back to discuss the shortcomings
of this approach in the next section. The main quantity calculated
through the Boltzmann equation is the non-equilibrium electronic distribution
function $f_{\mathbf{k}}$. In the linearized approximation, which
is valid for small departure from equilibrium, it can be expanded
as $f_{\mathbf{k}}=f_{\mathbf{k}}^{0}-\Phi_{\mathbf{k}}(\partial f_{\mathbf{k}}^{0}/\partial\varepsilon_{\mathbf{k}})$,
where $f_{\mathbf{k}}^{0}\equiv(e^{\beta\xi_{\mathbf{k}}}+1)^{-1}$
is the equilibrium Fermi-Dirac distribution. In this case, the Boltzmann
equation is a linear integral equation for $\Phi_{\mathbf{k}}$: 
\begin{equation}
-e\mathbf{E}\cdot\mathbf{v}_{\mathbf{k}}\left(\frac{\partial f_{\mathbf{k}}^{0}}{\partial\varepsilon_{\mathbf{k}}}\right)=\frac{1}{T}\sum\limits _{\mathbf{k}'}^ {}(\Phi_{\mathbf{k}}-\Phi_{\mathbf{k}'})f_{\mathbf{k}}^{0}(1-f_{\mathbf{k}'}^{0})t_{\mathbf{k},\mathbf{k}'},\label{Eq_LinearBE}
\end{equation}
where $\mathbf{E}$ represents a uniform electric field applied to
the electronic quasi-particles with charge $-e$ and $v_{\mathbf{k}}=\nabla_{\mathbf{k}}\varepsilon_{\mathbf{k}}$
is the momentum dependent velocity. The quantity $t_{\mathbf{k},\mathbf{k}'}$
is the collision integral; in our problem, we consider that momentum
relaxation is provided by collision with impurities. Using the Kadanoff-Baym
\textit{ansatz} \cite{Kadanoff-WAB(1962)} with the effective self-energy correction in Fig.
\ref{One_Loop}(a) and the Born approximation, the collision integral
evaluates to 
\begin{align}
t_{\mathbf{k},\mathbf{k}'}= & \frac{2g_{\text{imp}}^{2}}{\nu_{0}}\delta(\varepsilon_{\mathbf{k}}-\varepsilon_{\mathbf{k}'})+\frac{2g_{\text{nem}}^{2}}{\nu_{0}}h_{(\mathbf{k}+\mathbf{k}')/2}^{2}n(\varepsilon_{\mathbf{k}'}-\varepsilon_{\mathbf{k}})\nonumber \\
 & \times\operatorname{Im}[\chi_{\text{nem}}(\mathbf{k}'-\mathbf{k},\varepsilon_{\mathbf{k}'}-\varepsilon_{\mathbf{k}}+i\eta)],
\end{align}
where $\eta\rightarrow 0^{+}$, $g_{\text{imp}}$ and $g_{\text{nem}}$ [defined previously
in Eq. \eqref{Eq_Nem_Elec}] correspond, respectively, to the impurity
and nematic transition rate amplitudes, $n(\omega)\equiv (e^{\beta\omega}-1)^{-1}$ is the Bose-Einstein
distribution, and $\chi_{\text{nem}}(\mathbf{q},i\Omega_{n})$ is
the nematic susceptibility renormalized by the coupling to both elastic
and electronic degrees of freedom. Note that $g_{\mathrm{imp}}$ is
proportional to both the impurity scattering potential and the impurity
concentration. It is important to note that the nematic fluctuations
here are assumed to effectively behave as a bath that is always in
equilibrium. This can only be justified if there are additional processes
by which the nematic fluctuations equilibrate much faster than the
electronic ones. One option is the phonon subsystem, particularly
due to the coupling between the nematic order parameter and acoustic
phonons, as described by Eq. \eqref{Eq_Nem_Elec}. Morever, if nematic fluctuations
arise from separate degrees of freedom (e.g. composite spin order
parameters \cite{Fernandes-NP(2014)}), these fast equilibration processes can take place within
the nematic subsystem. However, if the nematic fluctuations arise
from a Pomeranchuk-like interaction between low-energy fermions, additional
bands are necessary to ensure a finite resistivity and avoid special
geometric cancellations \cite{Maslov-PRL(2011),Pal-LJPTS(2012),Wang-PRB(2019)}.

Generally, we can write the renormalized nematic susceptibility as:
\begin{align}
\chi_{\mathrm{nem}}^{-1}\left(\mathbf{q},i\Omega_{n}\right)&= \nu^{-1}_0\bigg[r_{0}+q^{2}+\frac{\Omega_{n}^{2}}{c^{2}}\nonumber\\
 & -\nu_0\Pi_{\mathrm{latt}}(\mathbf{q},i\Omega_{n})-\nu_0\Pi_{\mathrm{elec}}(\mathbf{q},i\Omega_{n})\bigg],\label{Eq_Chi_Nem}
\end{align}
where $\Pi_{\mathrm{latt}}(\mathbf{q},i\Omega_{n})$ and $\Pi_{\text{elec}}(\mathbf{q},i\Omega_{n})$
are bosonic self-energy corrections due to the coupling to elastic
fluctuations and particle-hole excitations {[}see Fig. \ref{One_Loop}(b){]}.
The former was exactly computed in Eq. \eqref{Eq_Latt_Bubble},
and its main effect is to replace $r_{0}$ by the renormalized mass
$r\left(\varphi\right)$ given by Eq. \eqref{Eq_Effective_Mass}.
As for the latter, within a Hertz-Millis approach, its main effect
is to change the dynamics of the nematic fluctuations by giving rise
to additional frequency-dependent terms: 
\begin{align}
\Pi_{\text{elec}}(\mathbf{q},i\Omega_{n})=&-g_{\text{nem}}^{2}\cos^{2}(2\varphi)\frac{|\Omega_{n}|}{v_{F}q}\nonumber \\
 & -4g_{\text{nem}}^{2}\sin^{2}(2\varphi)\biggl(\frac{\Omega_{n}}{v_{F}q}\biggr)^{2},\label{Eq_Pol_Bubble}
\end{align}
where $v_{F}\equiv|\mathbf{v}_{\mathbf{k}_{F}}|$ is the Fermi velocity.
Although the second term yields a sub-leading frequency dependence
as compared to the first term, it is the only non-zero term along
the directions $\varphi_{n}=(2n+1)\pi/4$ (see also Refs. \cite{Hartnoll-PRB(2014),Zacharias-PRB(2009),Paul-PRL(2017)}).
These are the same directions along which the renormalized mass $r\left(\varphi\right)$
vanishes, and along which the nematic form factor $h_{\mathbf{k}}$
vanishes, i.e. the cold spots directions.

Within the Boltzmann-equation formalism, the electrical resistivity
$\rho(T)$ can be computed by the minimization of the functional \cite{Hlubina-PRB(1995),Rosch-PRL(1999)}
\begin{equation}
\rho[\Phi]=\frac{1}{\nu_{0}e^{2}}\frac{\oint\oint\frac{d\mathbf{k}}{|\mathbf{v}_{\mathbf{k}}|}\frac{d\mathbf{k}'}{|\mathbf{v}_{\mathbf{k}'}|}\mathcal{F}_{\mathbf{k},\mathbf{k}'}(\Phi_{\mathbf{k}}-\Phi_{\mathbf{k}'})^{2}}{\Bigl[\oint\frac{d\mathbf{k}}{|\mathbf{v}_{\mathbf{k}}|}(\mathbf{v}_{\mathbf{k}}\cdot\hat{\bm{n}})\Phi_{\mathbf{k}}\Bigr]^{2}},\label{Eq_Func_Res}
\end{equation}
where the unitary vector $\hat{\bm{n}}$ points in the direction of
the electric field $\mathbf{E}$, the momentum integrals are defined
around the Fermi surface, and the function $\mathcal{F}_{\mathbf{k},\mathbf{k}'}$
encodes the scattering by impurities and nematic fluctuations. In
our case, it is given by 
\begin{align}
\mathcal{F}_{\mathbf{k},\mathbf{k}'}= & \;g_{\text{imp}}^{2}+g_{\text{nem}}^{2}\frac{h_{(\mathbf{k}+\mathbf{k}')/2}^{2}}{T}\int_{-\infty}^{\infty}d\omega\,\omega\,n(\omega)[n(\omega)+1]\nonumber \\
 & \times\operatorname{Im}[\chi_{\text{nem}}(\mathbf{k}-\mathbf{k}',\omega+i\eta)],\label{aux_F}
\end{align}
with the nematic susceptibility:
\begin{align}
\chi_{\mathrm{nem}}^{-1}\left(\mathbf{q},i\Omega_{n}\right)&= \nu^{-1}_0\bigg[r_{0}-r_{0,c}+\lambda_\text{latt}\cos^{2}(2\varphi)+q^{2}\nonumber\\
 & +\nu_0g_{\text{nem}}^{2}\frac{|\Omega_{n}|}{v_{F}q}\left(\cos^{2}2\varphi+4\sin^{2}2\varphi\,\frac{|\Omega_{n}|}{v_{F}q}\right)\bigg].\label{chi_nem_final}
\end{align}

Eq. \eqref{aux_F} can be evaluated analytically by using the residue
theorem and by linearizing the form factor $h_{\mathbf{k}}$ at the
Fermi surface. As a result, we find:
\begin{widetext}
\begin{align}
\mathcal{F}(\mathbf{q})= & \;g_{\text{imp}}^{2}+g_{\text{nem}}^{2}\nu_0\cos^{2}(2\varphi)\Biggl[\left(\frac{B_{\mathbf{q}}-\sqrt{B_{\mathbf{q}}^{2}-4A_{\mathbf{q}}R_{\mathbf{q}}}}{2\pi TA_{\mathbf{q}}\sqrt{B_{\mathbf{q}}^{2}-4A_{\mathbf{q}}R_{\mathbf{q}}}}\right)\psi^{(1)}\left(\frac{B_{\mathbf{q}}-\sqrt{B_{\mathbf{q}}^{2}-4A_{\mathbf{q}}R_{\mathbf{q}}}}{4\pi TA_{\mathbf{q}}}\right)-\left(\frac{B_{\mathbf{q}}+\sqrt{B_{\mathbf{q}}^{2}-4A_{\mathbf{q}}R_{\mathbf{q}}}}{2\pi TA_{\mathbf{q}}\sqrt{B_{\mathbf{q}}^{2}-4A_{\mathbf{q}}R_{\mathbf{q}}}}\right)\nonumber \\
 & \times\psi^{(1)}\left(\frac{B_{\mathbf{q}}+\sqrt{B_{\mathbf{q}}^{2}-4A_{\mathbf{q}}R_{\mathbf{q}}}}{4\pi TA_{\mathbf{q}}}\right)-\frac{2\pi T}{R_{\mathbf{q}}}\Biggr],\label{Eq_F}
\end{align}
where $\mathcal{F}_{\mathbf{k},\mathbf{k}'}=\mathcal{F}(\mathbf{k}-\mathbf{k}')$,
$\psi^{(1)}(z)$ is the trigamma function, $R_{\mathbf{q}}\equiv r_{0}-r_{0,c}+\lambda_\text{latt}\cos^{2}(2\varphi)+q^{2}$,
$A_{\mathbf{q}}\equiv4\nu_0g_{\text{nem}}^{2}\frac{\sin^{2}(2\varphi)}{(v_{F}q)^{2}}$,
and $B_{\mathbf{q}}\equiv \nu_0g_{\text{nem}}^{2}\frac{\cos^{2}(2\varphi)}{v_{F}q}$.
In order to simplify some of the numerical calculations, we will henceforth
approximate $\psi^{(1)}(z)$ by the simpler function 
\begin{equation}
\psi^{(1)}(z)\approx\frac{1}{z^{2}}+\frac{18z(z+1)(2z+1)}{(6z^{2}+6z+1)^{2}},\label{Eq_ApprTrigamma}
\end{equation}
which describes almost exactly the behavior of $\psi^{(1)}(z)$ for
$|z|>1$ \cite{Mortici-AMC(2010)}. We also note that this approximation does not change the
sign of $\mathcal{F}(\mathbf{q})$, which stays positive for all values
of the momentum $\mathbf{q}$ and the parameters of the model.
\end{widetext}

\begin{figure*}[t]
\centering \includegraphics[width=0.328\linewidth]{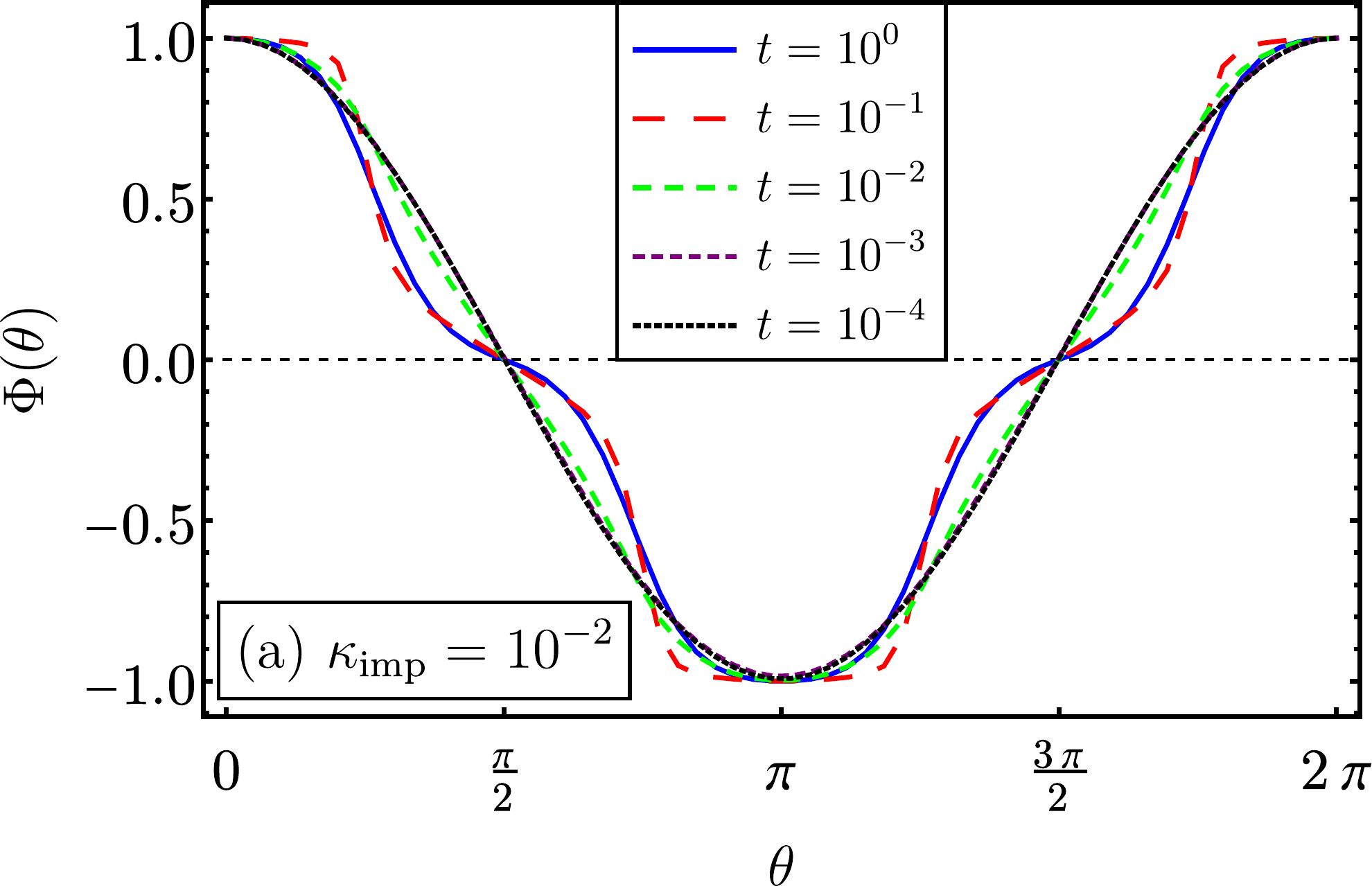}\hfill{}\includegraphics[width=0.328\linewidth]{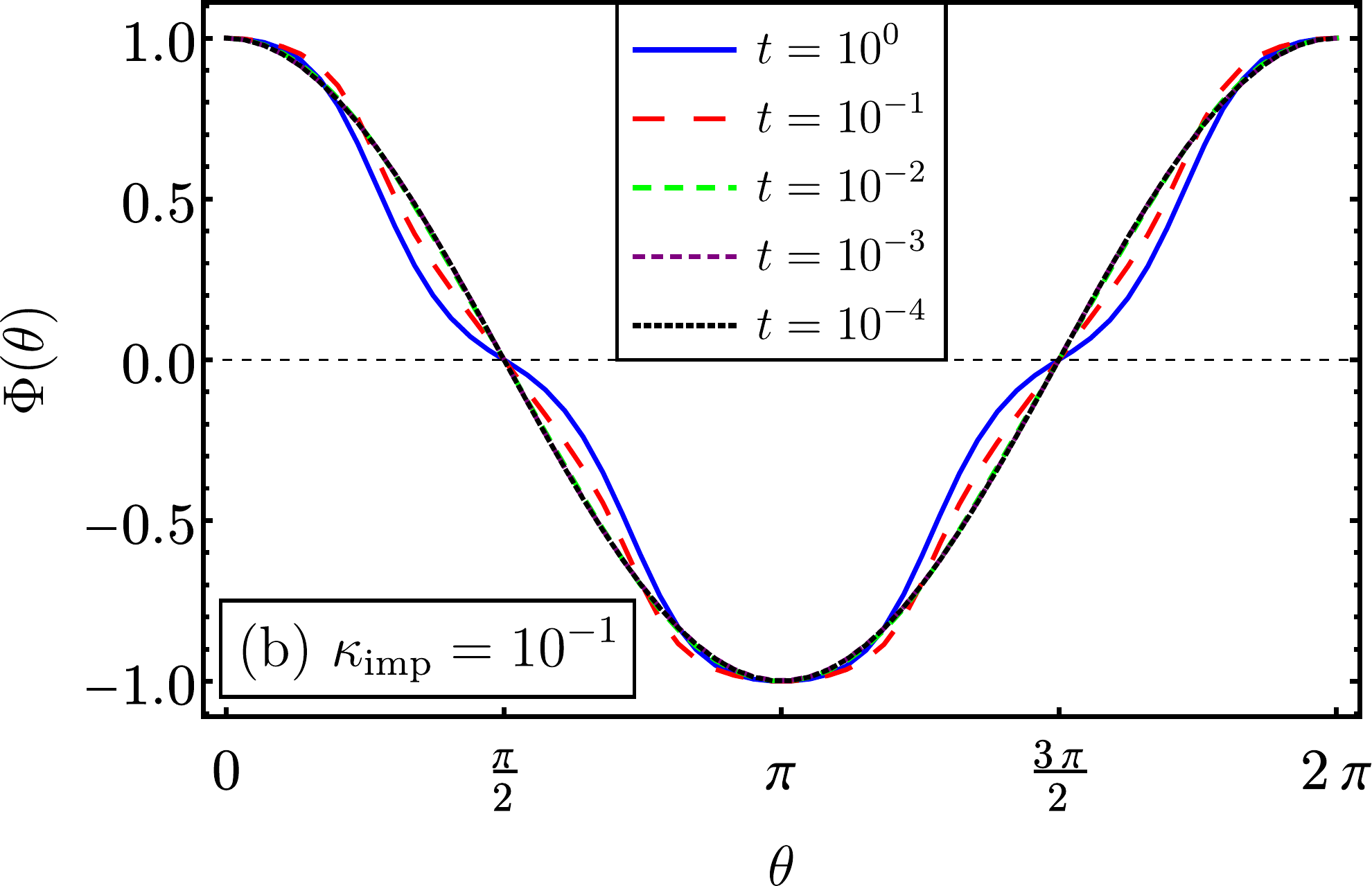}\hfill{}\includegraphics[width=0.328\linewidth]{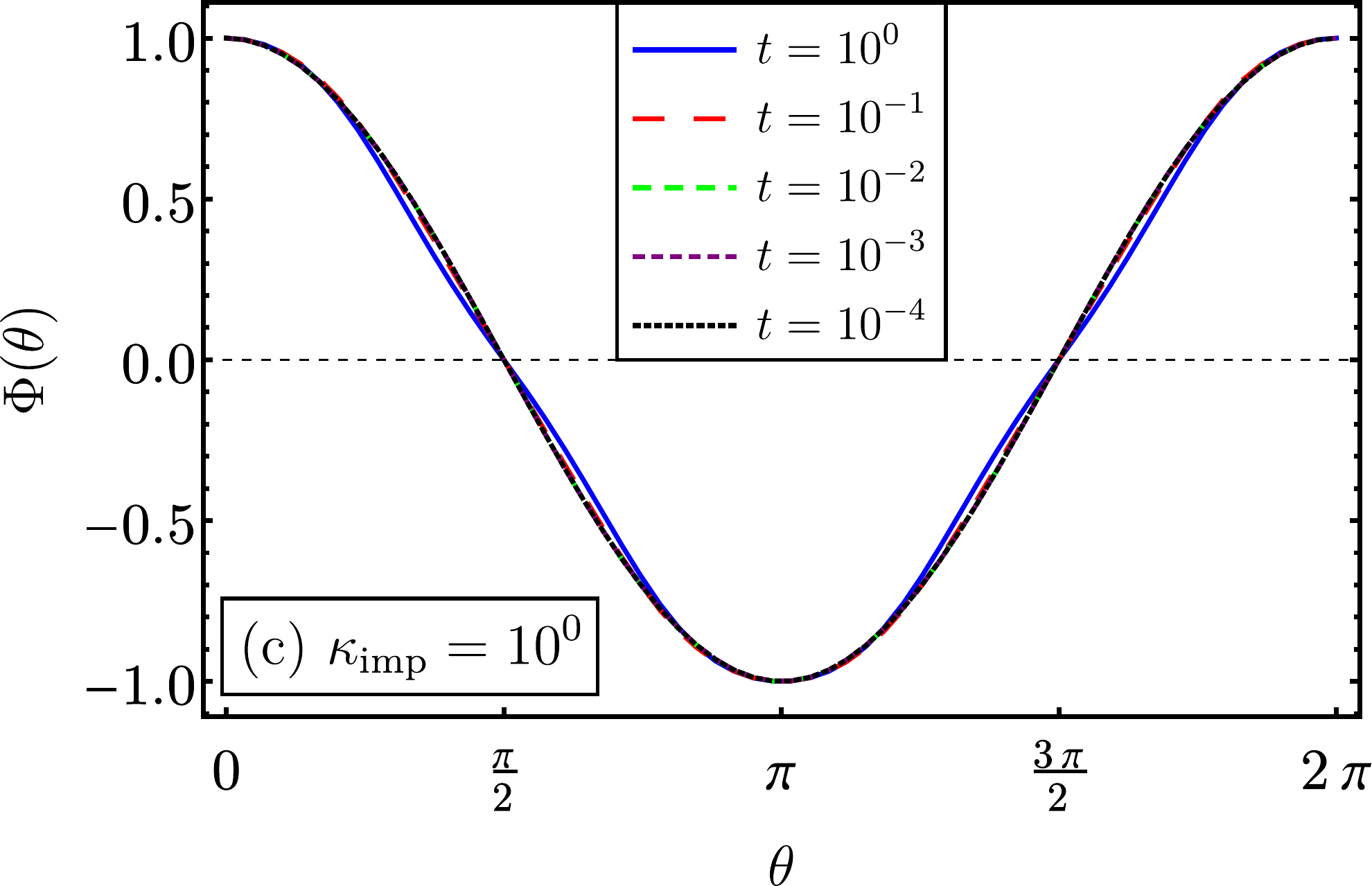}

\vfill{}
\includegraphics[width=0.328\linewidth]{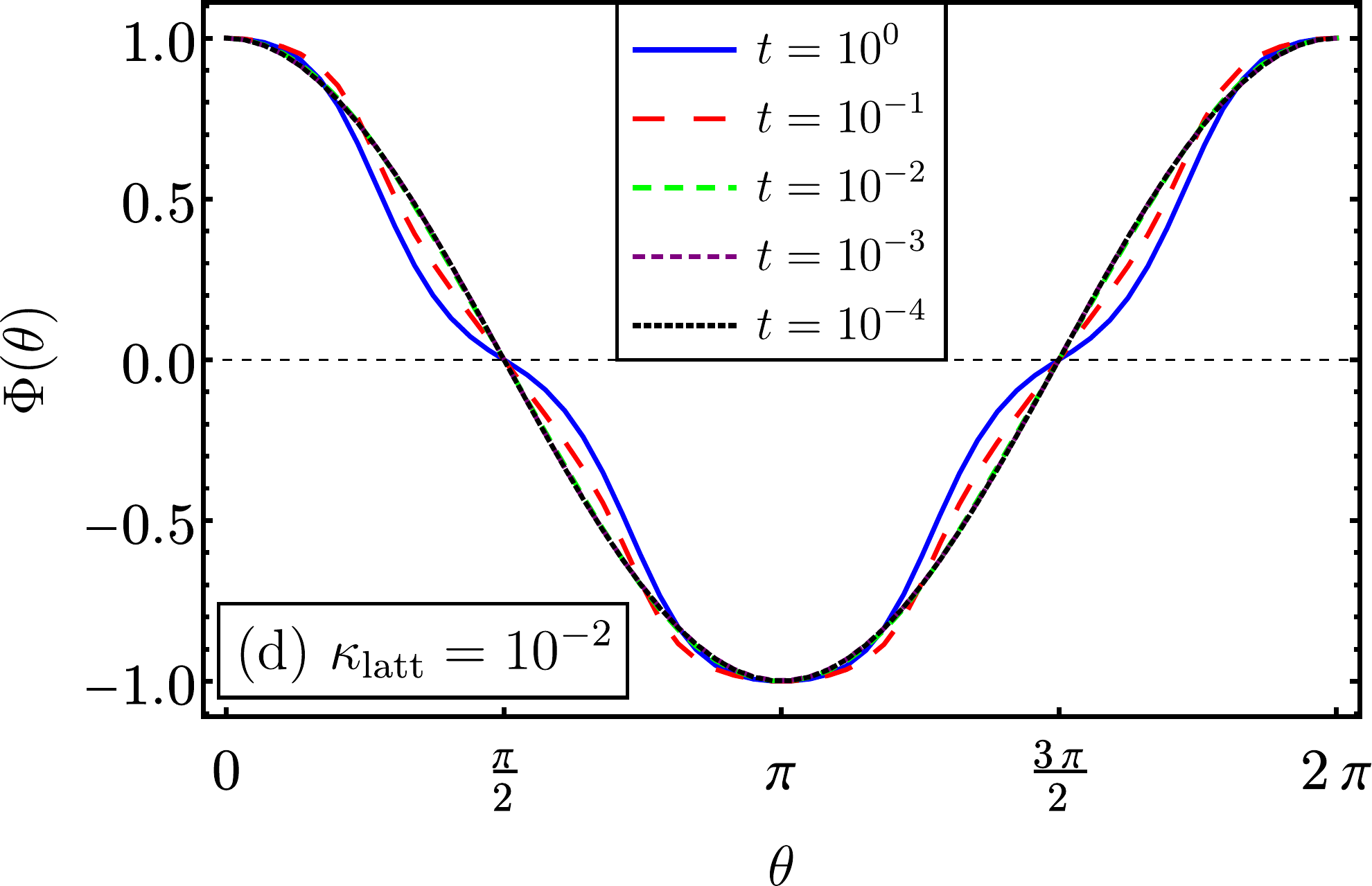}\hfill{}\includegraphics[width=0.328\linewidth]{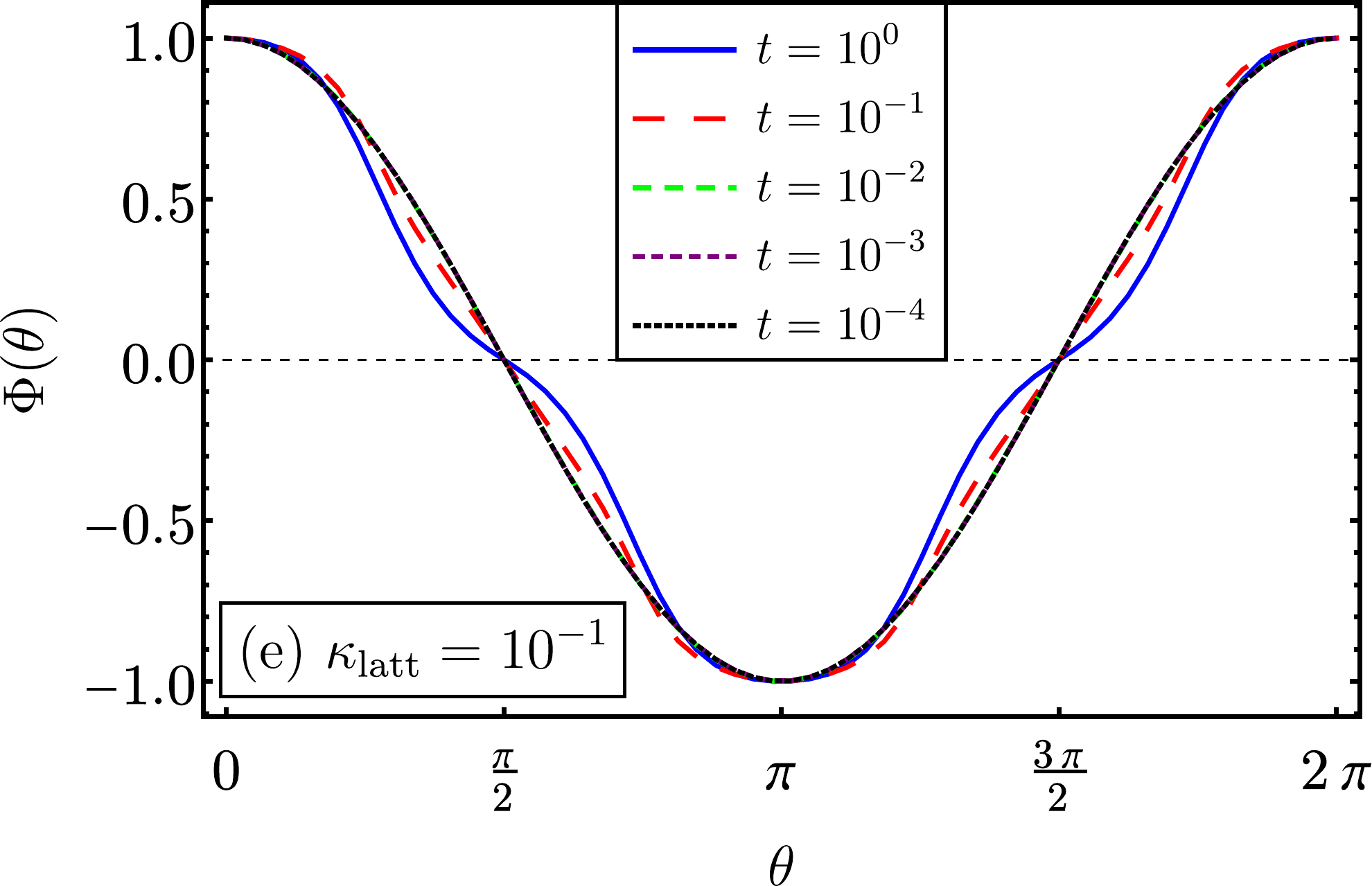}\hfill{}\includegraphics[width=0.328\linewidth]{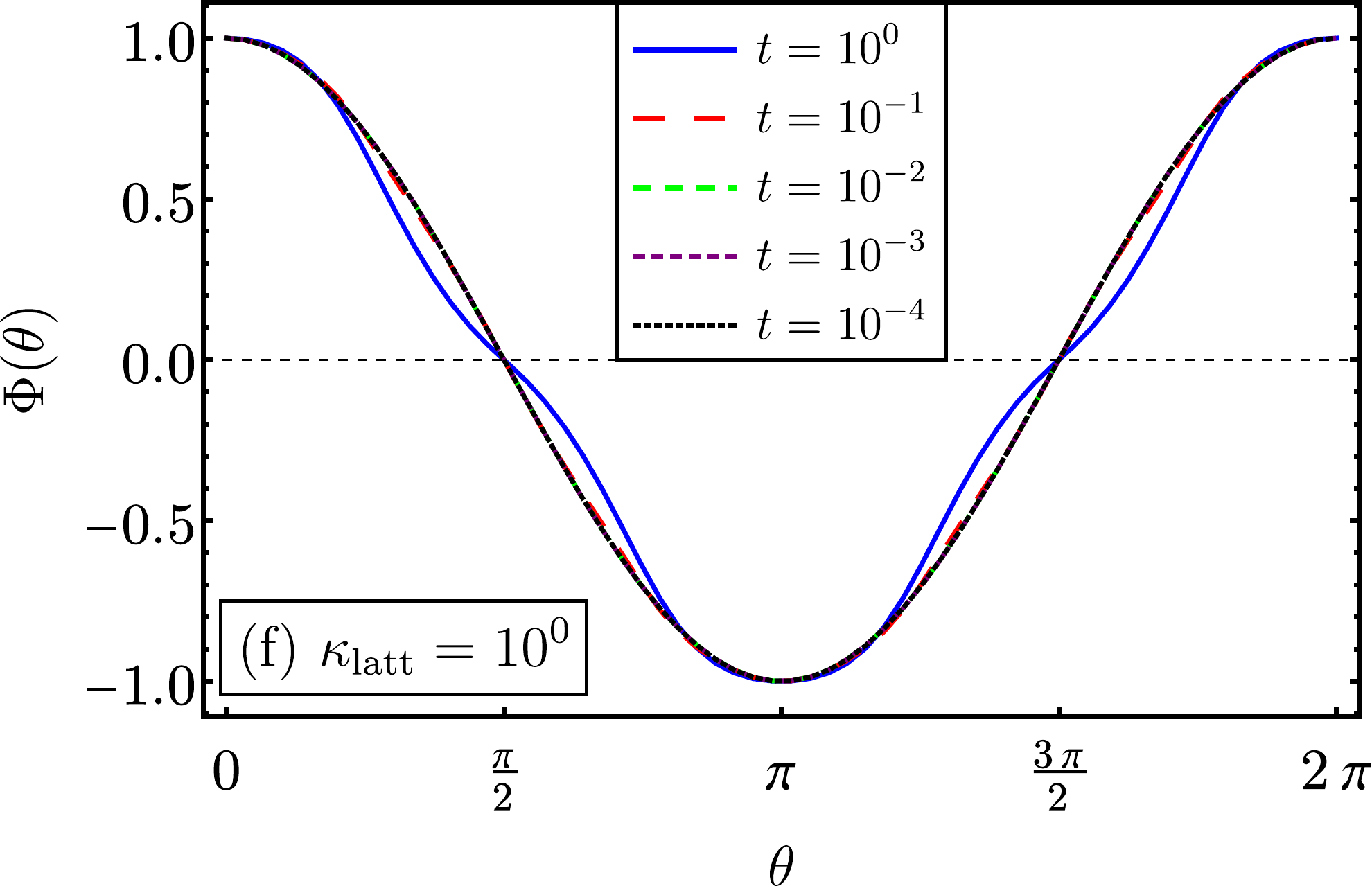}

\caption{(Color online) Behavior of the normalized quasi-particle distribution
$\Phi(\theta)$ obtained by solving Eq. \eqref{Eq_Fredholm} as a function of the reduced temperature $t=T/\varepsilon_{F}$,
the dimensionless impurity coupling $\kappa_\text{imp}=g^2_\text{imp}/g^2_\text{nem}$, and the dimensionless lattice coupling
$\kappa_\text{latt}=\lambda_\text{latt}/(\nu_0g^2_\text{nem})$.
In panels (a)\textendash (c), we fix $\kappa_\text{latt}=10^{-2}$
and vary $\kappa_\text{imp}$, whereas in panels (d)\textendash (f) we set $\kappa_\text{imp}=10^{-1}$
and then change the value of $\kappa_\text{latt}$. Notice that the effect on the quasi-particle distribution $\Phi(\theta)$ 
of increasing either the impurity coupling or the lattice
coupling is quite similar.
In addition, note that $\Phi(\theta)$ approaches asymptotically
the cosine function in the low-temperature limit.}
\label{Sol_Quasi_Distribution} 
\end{figure*}

The minimization of the functional $\rho[\Phi]$ or, in other words,
the demand of the condition $\delta\rho[\Phi]/\delta\Phi=0$, leads
to an integral equation for $\Phi_{\mathbf{k}}$ equivalent to the
one obtained by the integration of the momentum component perpendicular
to the Fermi surface in the Boltzmann equation itself {[}see Eq. \eqref{Eq_LinearBE}{]}.
For a circular Fermi surface, the equation for the distribution function
$\Phi$ in terms of the angle $\theta$ between the momentum $\mathbf{k}$
and the electric field $\mathbf{E}$ becomes a Fredholm equation of
the second kind given by
\begin{equation}
\Phi(\theta)=\int_{0}^{2\pi}d\theta'\mathcal{K}(\theta,\theta')\Phi(\theta')+f(\theta),\label{Eq_Fredholm}
\end{equation}
with:
\begin{align}
\mathcal{K}(\theta,\theta') & \equiv\dfrac{\mathcal{F}(\theta,\theta')}{\int_{0}^{2\pi}d\theta''\mathcal{F}(\theta,\theta'')},\\
f(\theta) & \equiv\dfrac{\zeta\cos(\theta)}{\int_{0}^{2\pi}d\theta''\mathcal{F}(\theta,\theta'')}.
\end{align}
Here, $\zeta=2\pi^{2}ev_{F}^{2}/k_{F}$ and $\mathcal{F}(\theta,\theta')$
is obtained by substituting $2\varphi=\theta+\theta'-\pi$
and $q^{2}=4k_{F}^{2}\sin^{2}[(\theta-\theta')/2]$ in Eq. \eqref{aux_F} for
$\mathcal{F}(\mathbf{q})$. It is convenient
to separate the terms arising from scattering by impurities and by
nematic fluctuations according to $\mathcal{F}(\theta,\theta')\equiv g_{\text{imp}}^{2}+g_{\text{nem}}^{2}\mathcal{F}_{\text{nem}}(\theta,\theta')$,
and express the resistivity as a sum of two terms:
\begin{equation}
\rho(T)=\rho_{\text{imp}}(T)+\rho_{\text{nem}}(T),\label{Eq_ResA}
\end{equation}
where 
\begin{align}
\rho_{\text{imp}}(T)=\frac{\kappa_{\mathrm{imp}}\rho^0_{\text{nem}}}{4}\frac{\int_{0}^{2\pi}\int_{0}^{2\pi}d\theta d\theta'[\Phi(\theta)-\Phi(\theta')]^{2}}{\Bigl[\int_{0}^{2\pi}d\theta\cos(\theta)\Phi(\theta)\Bigr]^{2}},\label{Eq_ResB}
\end{align}
\begin{align}
\rho_{\text{nem}}(T)=\frac{\rho^0_{\text{nem}}}{4}\frac{\int_{0}^{2\pi}\int_{0}^{2\pi}d\theta d\theta'\mathcal{F}_{\text{nem}}(\theta,\theta')[\Phi(\theta)-\Phi(\theta')]^{2}}{\Bigl[\int_{0}^{2\pi}d\theta\cos(\theta)\Phi(\theta)\Bigr]^{2}}.\label{Eq_ResC}
\end{align}

For convenience, we defined the ratio between the impurity and nematic
transition rate amplitudes, $\kappa_{\mathrm{imp}}\equiv g_{\text{imp}}^{2}/g_{\text{nem}}^{2}$, which we refer hereafter as the dimensionless impurity coupling,
and the nematic resistivity scale $\rho^0_{\text{nem}}\equiv4g_{\text{nem}}^{2}/(\nu_{0}e^{2}v_{F}^{2})$.
Notice that the residual resistivity is given simply by $\rho_{0}\equiv\kappa_{\mathrm{imp}}\rho^0_{\mathrm{nem}}$.
For later convenience, we also define the ratio between the lattice
coupling and the nematic coupling, $\kappa_{\mathrm{latt}}\equiv\lambda_\text{latt}/(\nu_0g_{\text{nem}}^{2})=\frac{1}{2C_{66}}\left(1+\frac{C_{12}}{C_{11}}\right)\frac{g_{\mathrm{latt}}^{2}}{g_{\text{nem}}^{2}}$, which we refer hereafter as the dimensionless lattice coupling,
and the reduced temperature $t\equiv T/\varepsilon_{F}$, where $\varepsilon_{F}=v_F k_F$ denotes the Fermi energy.

\subsection{Solution of the Boltzmann equation}

\begin{figure*}[t]
\centering \includegraphics[width=0.332\linewidth]{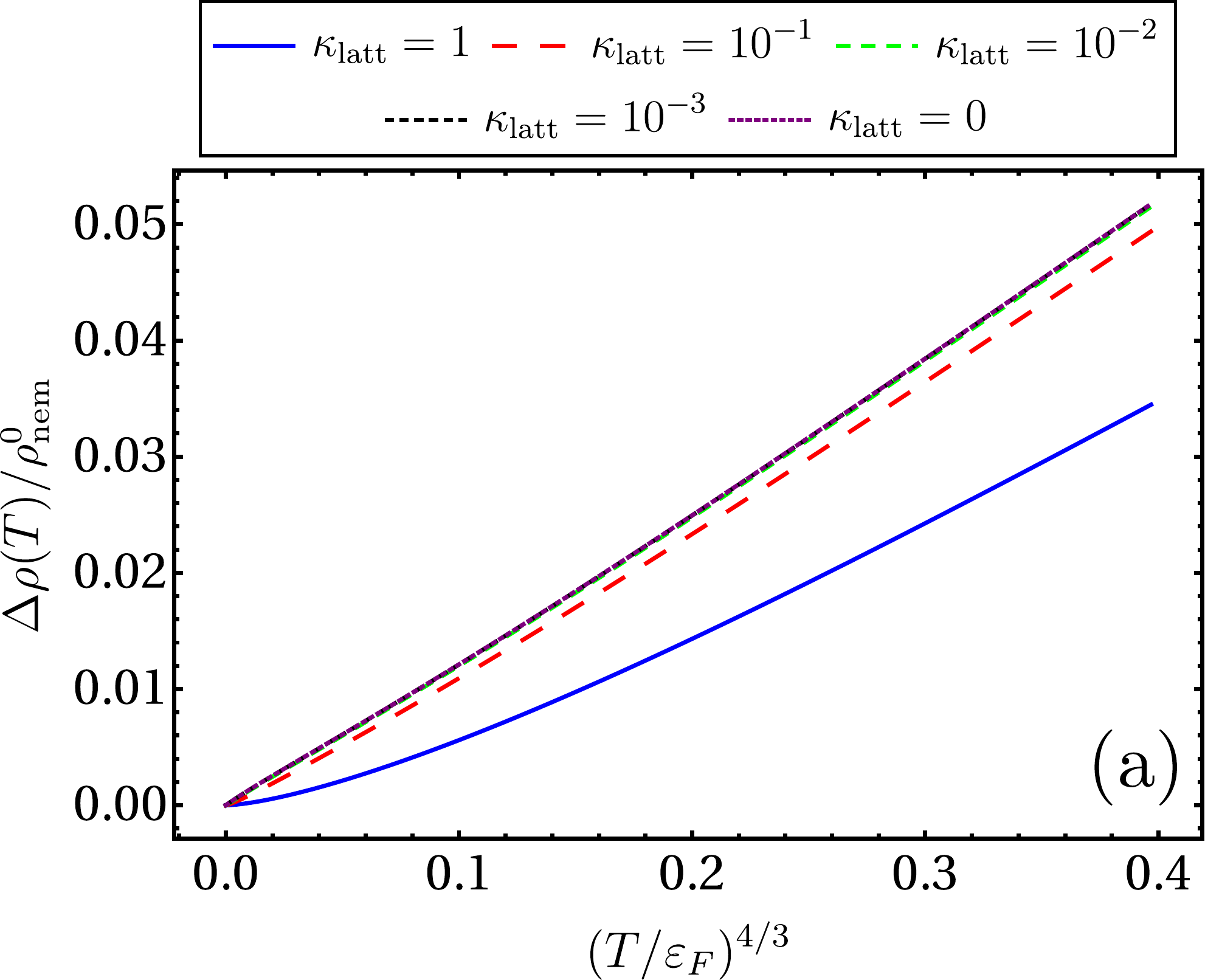}\hfill{}\includegraphics[width=0.332\linewidth]{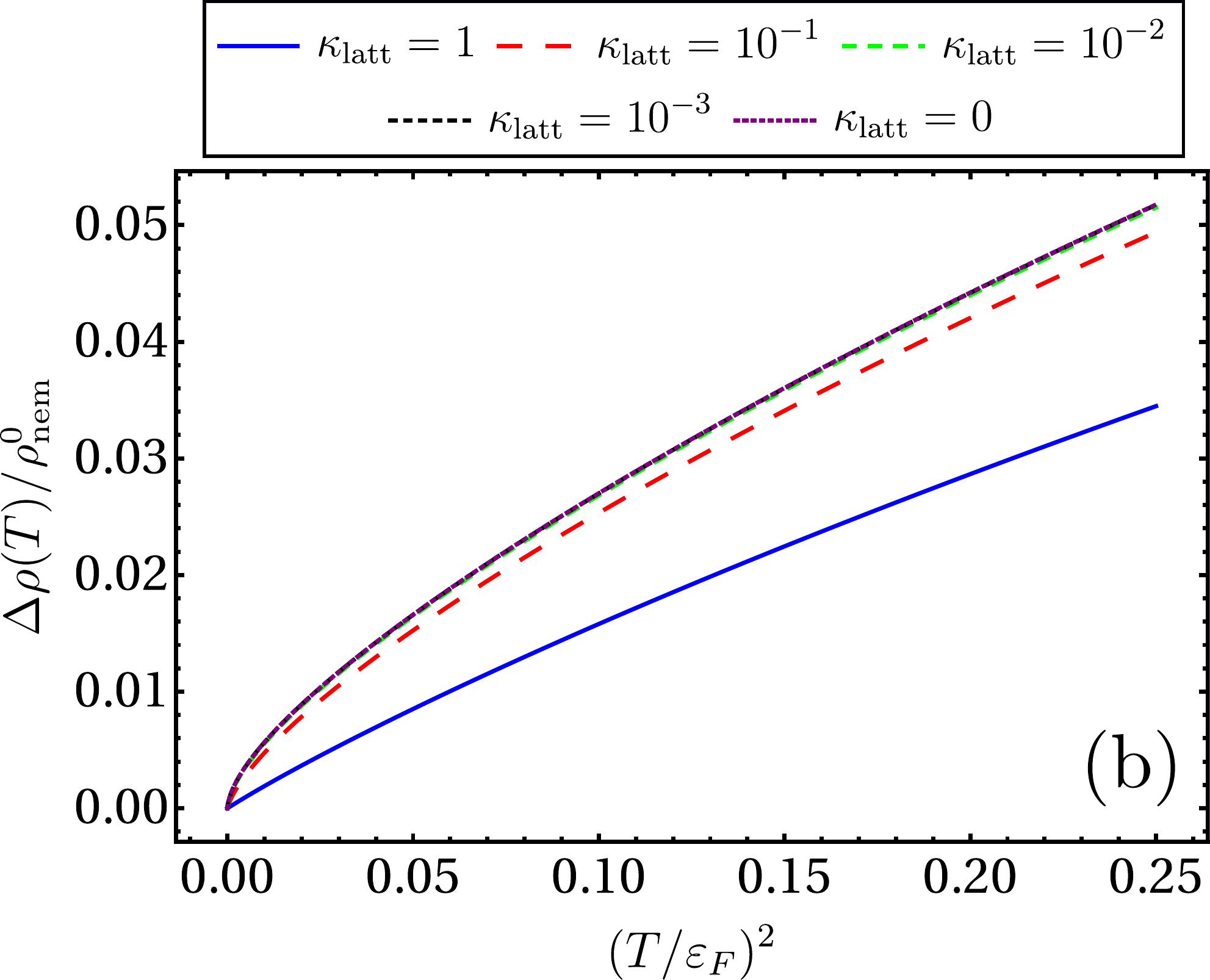}\hfill{}\includegraphics[width=0.330\linewidth]{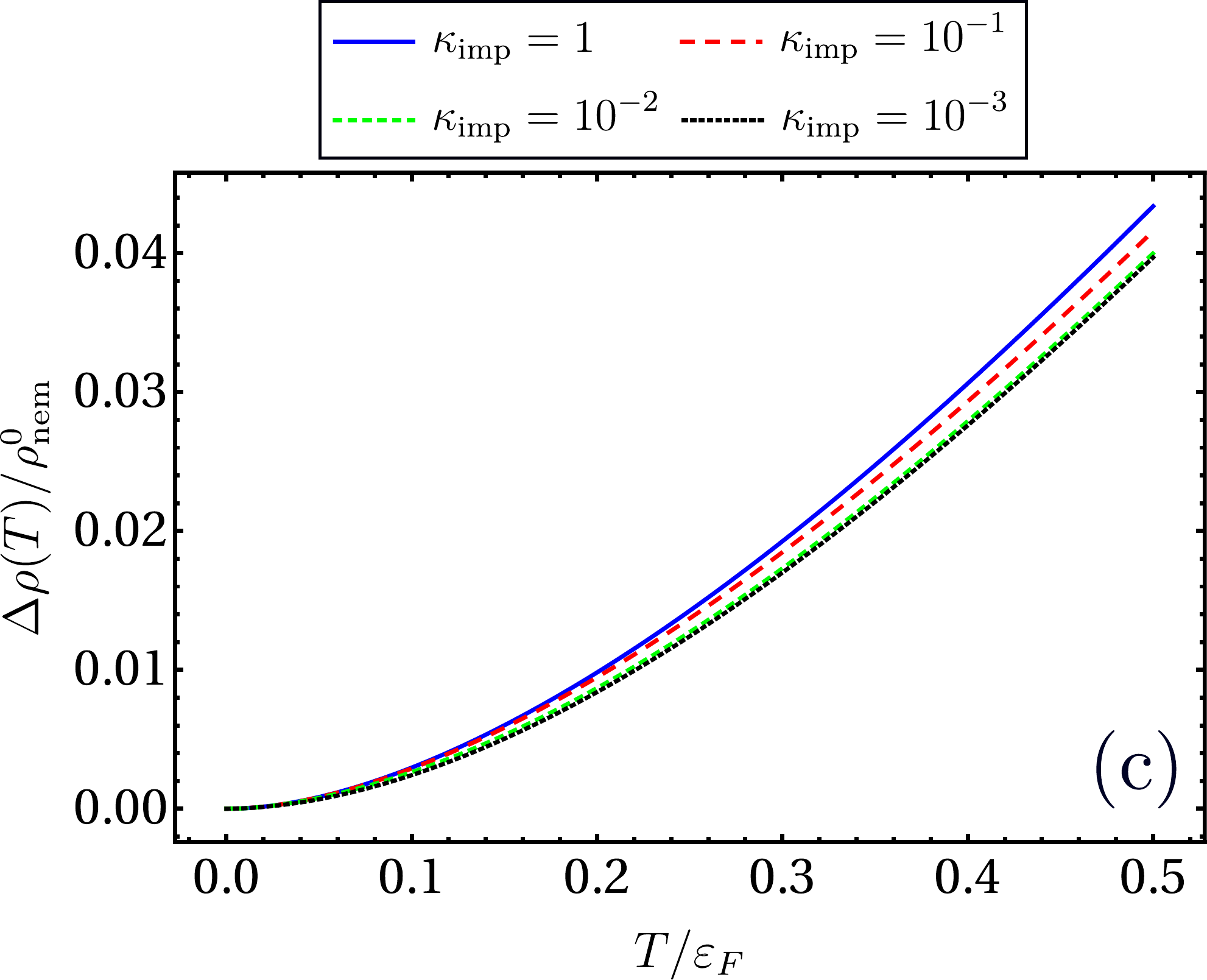}

\caption{(Color online) (a-b) Behavior of the resistivity $\Delta\rho(T)=\rho(T)-\rho_0$
as a function of the temperature $T$ and the dimensionless lattice coupling
$\kappa_\text{latt}$. We fix here the dimensionless impurity coupling to $\kappa_\text{imp}=10^{-1}$. In the absence of
$\kappa_\text{latt}$, the low-temperature behavior of the resistivity is
characterized by $\Delta\rho(T)\sim T^{4/3}$ {[}panel (a){]},
although when $\kappa_\text{latt}$ is $\mathcal{O}(1)$, namely when the lattice coupling $g_\text{latt}$ is comparable to the nematic interaction $g_\text{nem}$, the resistivity is given by $\Delta\rho(T)\sim T^{2}$ {[}panel (b){]}. (c) This behavior is rather insensitive to changes in the impurity coupling $\kappa_\text{imp}$, particularly at low temperatures. In this particular case, we fix $\kappa_\text{latt}=0.5$.}
\label{Sol_Resistivity} 
\end{figure*}

We first solve analytically the Boltzmann equation in the low-temperature
regime near the QCP, which corresponds to setting $r_{0}=r_{0,c}$.
At low enough temperatures, inelastic scattering by nematic fluctuations
is always subleading with respect to the elastic scattering by impurities.
Of course, this temperature scale depends on the dimensionless lattice
coupling $\kappa_{\mathrm{latt}}$, which we consider to be finite.
In this regime, the out-of-equilibrium distribution function can be
well approximated by the solution of the Boltzmann equation in the
presence of impurity scattering only, which gives the standard expression
$\Phi(\theta)=\cos(\theta)$ for the normalized distribution function \cite{Ziman-OUP(1960)}.

We first consider the case in which the coupling to the lattice vanishes,
$\kappa_{\mathrm{latt}}=0$. By employing Eq. \eqref{Eq_ApprTrigamma}, we are able to approximate $\mathcal{F}(\mathbf{q})$ in the low-temperature limit by the expression
\begin{equation}
\mathcal{F}(\mathbf{q})\approx g^2_\text{imp}+g^2_\text{nem}\nu_0\cos^2(2\varphi)\frac{4\pi^2 B_\mathbf{q}T^2}{3(R_\mathbf{q}+2\pi B_\mathbf{q}T)^2},
\end{equation}
where $R_\mathbf{q}=r_0+q^2$. In this situation, the electrical resistivity [see Eqs. \eqref{Eq_ResA}--\eqref{Eq_ResC}] evaluates to
\begin{equation}\label{Eq_Exact_ResA}
\rho(T)=\rho^0_\text{nem}\bigg[\kappa_\text{imp}+\frac{2^{1/3}\Gamma\left(-\frac{4}{3}\right)\Gamma\left(\frac{11}{6}\right)\nu^{1/3}_0g^{2/3}_\text{nem}}{27\pi^{7/6}k^{2/3}_F}\bigg(\frac{T}{\varepsilon_F}\bigg)^{4/3}\bigg],
\end{equation}
with $\Gamma(z)$ being the Gamma function. Therefore, we reproduce within the Boltzmann-equation formalism the expected $T^{4/3}$ scaling behavior of the resistivity for a dirty 2D electronic system close to a nematic QCP.

We now move to the case where the coupling to the lattice $\kappa_\text{latt}$ is finite. By making use once again of Eq. \eqref{Eq_ApprTrigamma} and then considering the temperature range $T\ll\kappa^{3/2}_\text{latt}\varepsilon_F$, one finds
\begin{equation}
\mathcal{F}(\mathbf{q})\approx g^2_\text{imp}+g^2_\text{nem}\nu_0\cos^2(2\varphi)\frac{4\pi^2 B_\mathbf{q}T^2}{3R^2_\mathbf{q}}.
\end{equation}
As a result, for the range of nemato-elastic interactions defined by $0<\kappa_\text{latt}\lesssim k^2_F/(\nu_0g^2_\text{nem})$, the electrical resistivity at the nematic QCP becomes
\begin{equation}\label{Eq_Exact_ResB}
\rho(T)=\rho^0_\text{nem}\bigg[\kappa_\text{imp}+\frac{\mathcal{C}_0}{\kappa_\text{latt}}\bigg(\frac{T}{\varepsilon_F}\bigg)^2\bigg],
\end{equation}
where $\mathcal{C}_0\approx 0.17$ is a numerical constant. Consequently, the resistivity in this particular case describes Fermi-liquid-like behavior, which agrees with the results in Ref. \cite{Paul-PRL(2017)} based on the calculations of thermodynamics and single-particle properties.

Our asymptotic analysis for the low-temperature behavior of the resistivity
suggests that a crossover from $\Delta\rho(T)\sim T^{4/3}$ at moderate
temperatures to $\Delta\rho(T)\sim T^{2}$ at low temperatures can
be expected, with the crossover temperature scale being determined
by the dimensionless lattice coupling $\kappa_{\mathrm{latt}}$. To
verify this expectation, we numerically solve the integral equation
(\ref{Eq_Fredholm}) to find the non-equilibrium distribution function
$\Phi\left(\theta\right)$ and then compute the resistivity in Eqs.
(\ref{Eq_ResB}) and (\ref{Eq_ResC}). To make the numerical calculations
convergent, we consider that at the QCP the effective nematic mass
vanishes linearly with temperature, i.e. $r_{0}=r_{0,c}+a(T/\varepsilon_F)$. Such a linear $T$ dependence is characteristic of a mean-field behavior, which is theoretically expected to the be case for the nematic transition due to the coupling to the lattice (see Ref. \cite{Schmalian-PRB(2016)}); this is also the experimentally observed behavior \cite{Kuo-S(2016),Bohmer-JPCM(2018)}.
Here, we set the dimensionless constant $a$ to be $a=1$. 
As explained in more details in the Appendix, the low-temperature 
behavior of $\Delta\rho(T)$ is independent of the value of $a$.

In Fig. \ref{Sol_Quasi_Distribution}, we show the numerical solution
for the non-equilibrium distribution function $\Phi(\theta)$ by varying
the reduced temperature $T/\varepsilon_{F}$, the impurity coupling
$\kappa_{\mathrm{imp}}$, and the lattice coupling $\kappa_{\mathrm{latt}}$.
It is clear that, at low enough temperatures, the distribution function
always approaches the $\cos(\theta)$ function, characteristic of the
impurity-scattering only problem. As discussed above, this is a consequence
of the fact that, at low enough temperatures, inelastic scattering
is subleading compared to elastic scattering. As a result, by comparing
panels (a)-(c), which have the same $\kappa_{\mathrm{latt}}=10^{-2}$
parameter, it is clear that the temperature scale below which the
distribution function approaches $\cos(\theta)$ increases as $\kappa_{\mathrm{imp}}$
increases. Above this temperature scale, the main deviations from
the $\cos(\theta)$ distribution are located at the angles  $\theta_n=(2n+1)\pi/4$, which correspond to the cold spots of the Fermi
surface. Although this may seem contradictory at first sight, since
the nematic form factor vanishes for quasi-particle scattering at the cold spots, one can understand
this behavior as arising from the simple fact that we have to
average over all processes within the Fermi surface to find the quasi-particle distribution. Consequently, this also depends on the Fermi-surface regions where the nematic form-factor contribution is finite. Furthermore, we also notice that the zeros of the quasi-particle distribution $\Phi(\theta)$ occur at the same points where the function $\cos(\theta)$ goes to zero. In fact, at these points the Boltzmann equation becomes a homogeneous integral equation [see Eq. \eqref{Eq_Fredholm}], which has only a trivial solution due to the dependence of $\mathcal{K}(\theta,\theta')$ on the impurity coupling $\kappa_\text{imp}$.

In panels (d)--(f), we fix $\kappa_{\mathrm{imp}}=10^{-1}$ and vary
the dimensionless lattice coupling $\kappa_{\mathrm{latt}}$. It is interesting to note
that the nemato-elastic coupling plays a similar role as the disorder
coupling, in the sense that it also favors a distribution function
that is similar to the $\cos(\theta)$ distribution. The reason is because
the coupling to the lattice makes the nematic mass finite everywhere
except at the cold spots, thus removing much of the strongly anisotropic behavior
associated with the nematic QCP. 

Having determined $\Phi(\theta)$ numerically, we plot in Fig. \ref{Sol_Resistivity}
the temperature-dependence of the resistivity $\Delta\rho(T)\equiv\rho(T)-\rho_{0}$.
In panels (a)-(b), we fix the impurity coupling to $\kappa_{\mathrm{imp}}=10^{-1}$
and vary the lattice coupling $\kappa_{\mathrm{latt}}$. As
shown in panel (a), when the nematic degrees of freedom are uncoupled
from the lattice, $\kappa_{\mathrm{latt}}=0$, we find the expected
$\Delta\rho(T)\sim T^{4/3}$ behavior of a nematic QCP. Upon increasing
$\kappa_{\mathrm{latt}}$, we start noting deviations from
this power law. In particular, when $\kappa_{\mathrm{latt}}\sim1$,
there is a wide range of temperatures in which $\Delta\rho(T)\sim T^{2}$,
as illustrated in panel (b). However, for intermediate values of $\kappa_{\mathrm{latt}}$,
it is clear that the temperature dependence of the resistivity cannot
be described by a single power law. This is reminiscent of the effect
of impurity scattering on the resistivity near an antiferromagnetic
QCP, which makes the temperature dependence of $\Delta\rho(T)$ not display
a simple power-law behavior \cite{Rosch-PRL(1999),Rosch-PRB(2000)}. In our case, however, impurity scattering
has little effect on the temperature dependence of $\Delta\rho(T)$,
as shown in panel (c). The main effect comes from the coupling to
the lattice $\kappa_{\mathrm{latt}}$, which endows the nematic susceptibility
with an anisotropic correlation length.

\begin{figure}
\centering \includegraphics[width=0.95\columnwidth]{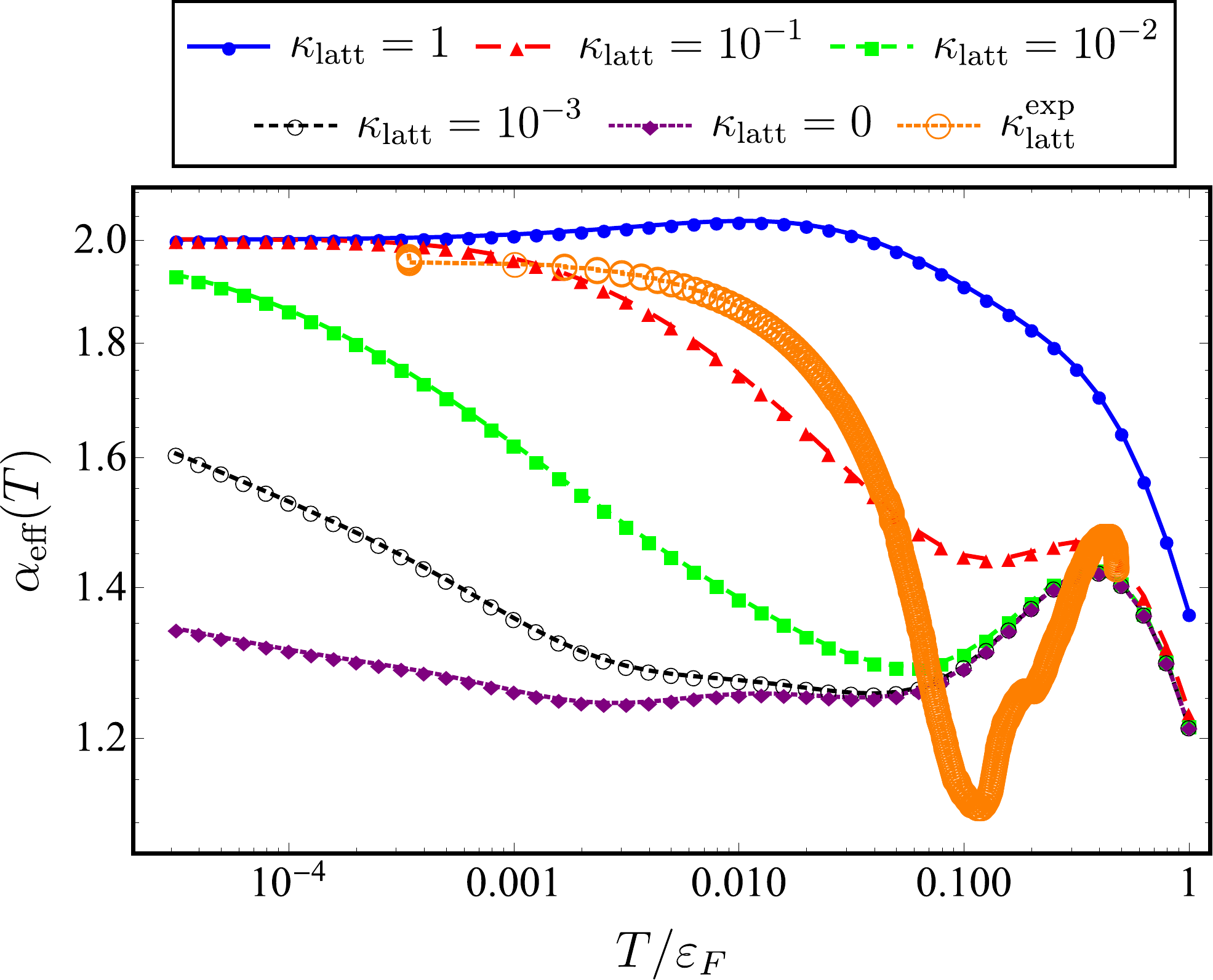}

\caption{(Color online) Temperature-dependent effective exponent $\alpha_\mathrm{eff}(T)$ of the electrical resistivity $\Delta\rho(T)\sim T^{\alpha_\mathrm{eff}(T)}$ for different values of the dimensionless lattice coupling $\kappa_\text{latt}$ and fixed dimensionless impurity coupling $\kappa_\text{imp}=10^{-1}$. At very low temperatures and at a finite value of $\kappa_\text{latt}$, $\alpha_\mathrm{eff}(T)$ will eventually saturate at $\alpha_\mathrm{eff}=2$, which marks the emergence of Fermi-liquid regime for the system. The bigger empty circles, indicated by $\kappa^\text{exp}_\text{latt}$, are the experimental data reported in Ref. \cite{Bristow2019} for the
resistivity exponent $\alpha_\mathrm{eff}(T)$ of ``optimally" doped FeSe$_{1-x}$S$_{x}$ with $x=0.18$, which harbors a putative nematic QCP. Here, we used the value $\varepsilon_F=250$ K for the Fermi energy of this material.}
\label{fig_exponent} 
\end{figure}

To better illustrate the behavior of $\Delta\rho(T)$ near the nematic
QCP coupled to the lattice, we plot in Fig. \ref{fig_exponent}
the effective temperature-dependent exponent $\alpha_{\mathrm{eff}}(T)\equiv\partial\ln\left[\Delta\rho(T)\right]/\partial\ln T$
for different values of the lattice coupling $\kappa_{\mathrm{latt}}$. It is clear that,
for $\kappa_{\mathrm{latt}}<1$, the exponent varies between the approximate
range $4/3\lesssim\alpha_{\mathrm{eff}}(T)\lesssim2$, as anticipated
from our analytical results. Note that the $T^{2}$ behavior may be
achieved only at extremely low temperatures, depending on the value
of $\kappa_{\mathrm{latt}}$. Similarly, the $T^{4/3}$ behavior may
be essentially inaccessible if $\kappa_{\mathrm{latt}}$ is not small
enough.

\section{Concluding remarks}

\label{Sec_IV}

In summary, we evaluated the impact of the coupling to the elastic
degrees of freedom on the electrical resistivity near a two-dimensional
metallic nematic QCP. Our main result is that the temperature-dependent
resistivity $\Delta\rho(T)=\rho(T)-\rho_{0}$ cannot generally be
described by a simple power law. This is a consequence of the fact
that the elastic coupling favors, at low enough temperatures, a Fermi-liquid
like behavior, characterized by $\Delta\rho(T)\sim T^{2}$. In contrast,
quantum critical nematic fluctuations, which are cutoff by the lattice
coupling, favor a $\Delta\rho(T)\sim T^{4/3}$ behavior. As a consequence
of these opposing tendencies, the effective exponent $\Delta\rho(T)\sim T^{\alpha_{\mathrm{eff}(T)}}$
shows a pronounced temperature dependence, as illustrated in Fig.
\ref{fig_exponent}, roughly crossing over between $4/3$ and $2$.

The lattice is always present in real systems, and its effects on
the nematic degrees of freedom are profound even in the qualitative
level, as manifested in the directional-dependence of the nematic
correlation length. Thus, a full understanding of experimental data
requires elucidating how the elastic degrees of freedom affect the
transport properties near the nematic QCP. Before discussing comparisons
with experimental results, it is important to further discuss the
limitations of our approach. The main reason to consider a Boltzmann-equation
approach is because quasi-particles are well-defined near the nematic
QCP due to the coupling to the elastic degrees of freedom, as previously
shown in Ref. \cite{Paul-PRL(2017)}. Of course, as the coupling to the lattice becomes
smaller, this approximation becomes more questionable. Furthermore,
even within this approximation, it is not obvious that the Hertz-Millis
approach employed here to account for the dynamics of the nematic
fluctuations will hold. It would be interesting, in this regard, to
go beyond the Boltzmann equation approach and consider a different
technique, such as the memory matrix formalism \cite{Forster-HFBSCF(1975)}. Previous applications
of this approach to the transport properties of the nematic QCP revealed
important deviations from the expectations of the Boltzmann-equation
approach under certain conditions \cite{Hartnoll-PRB(2014),Wang-PRB(2019)}. In particular, the recent work \cite{Wang-PRB(2019)} found an interesting broad temperature range in which the resistivity displays a linear-in-$T$ behavior, which is absent in the Boltzmann formalism. However, the impact of the lattice
degrees of freedom was not considered in those investigations. Similarly,
sign-problem-free Quantum Monte Carlo simulations \cite{Berg-PRX(2016),Lederer-PNAS(2017)} of the coupled nemato-elastic
QCP would be desirable.

The most transparent experimental evidence of an isolated putative nematic QCP is on
the phase diagram of FeSe$_{1-x}$S$_{x}$ \cite{Hosoi-PNAS(2016)}. While the very
large values of the nematic susceptibility and its temperature dependence
suggest a second-order transition, the sudden drop of the nematic transition temperature for a
small change of doping concentration $x$ is typical of first-order
transitions. The temperature dependence of the resistivity near this possible nematic QCP has been experimentally studied. Ref. \cite{Licciardello-N(2019)} finds a wide temperature
regime near the nematic QCP at the concentration $x_c$ where the resistivity displays a nearly linear behavior, which
is not captured by our model. This could be suggestive of additional excitations at play or non-Hertz-Millis behavior. On the other hand, Ref. \cite{Bristow2019} reported a temperature
dependence of $\alpha_{\mathrm{eff}}$ that is qualitatively consistent
with our findings, increasing from close to $3/2$ at higher temperatures
to close to $2$ at lower temperatures. The data points are also shown in Fig. \ref{fig_exponent}, for comparison. A similar temperature dependence of $\alpha_{\mathrm{eff}}$ was observed in Ref. \cite{Coldea-arXiv(2019)} when the nematic QCP was tuned by pressure in an ``underdoped'' composition of FeSe$_{1-x}$S$_{x}$. Of course, while our model is certainly
too simplified to capture the complex band structure of FeSe$_{1-x}$S$_{x}$,
the overall trend in $\alpha_{\mathrm{eff}}$ is what one would
expect from our analysis. Overall, our work unveils the crucial role played by the lattice on the transport properties near a nematic QCP.

\section*{Acknowledgments}

We would like to thank A. Chubukov, A. Coldea, A. Klein, E. Miranda,
H. Freire, I. Paul, and A. Schofield for stimulating discussions.
V.S.deC. thanks the financial support from FAPESP under Grant No.
2017/16911-3. R.M.F. is supported by the U.S. Department of Energy,
Office of Science, Basic Energy Sciences, under Award No. DE-SC0012336.



\setcounter{equation}{0}
\setcounter{figure}{0}
\renewcommand{\theequation}{A\arabic{equation}}
\renewcommand{\thefigure}{A\arabic{figure}}

\section*{Appendix: Details of the numerical solution of the Boltzmann equation}

\begin{figure}[t]
\centering \includegraphics[width=0.95\linewidth]{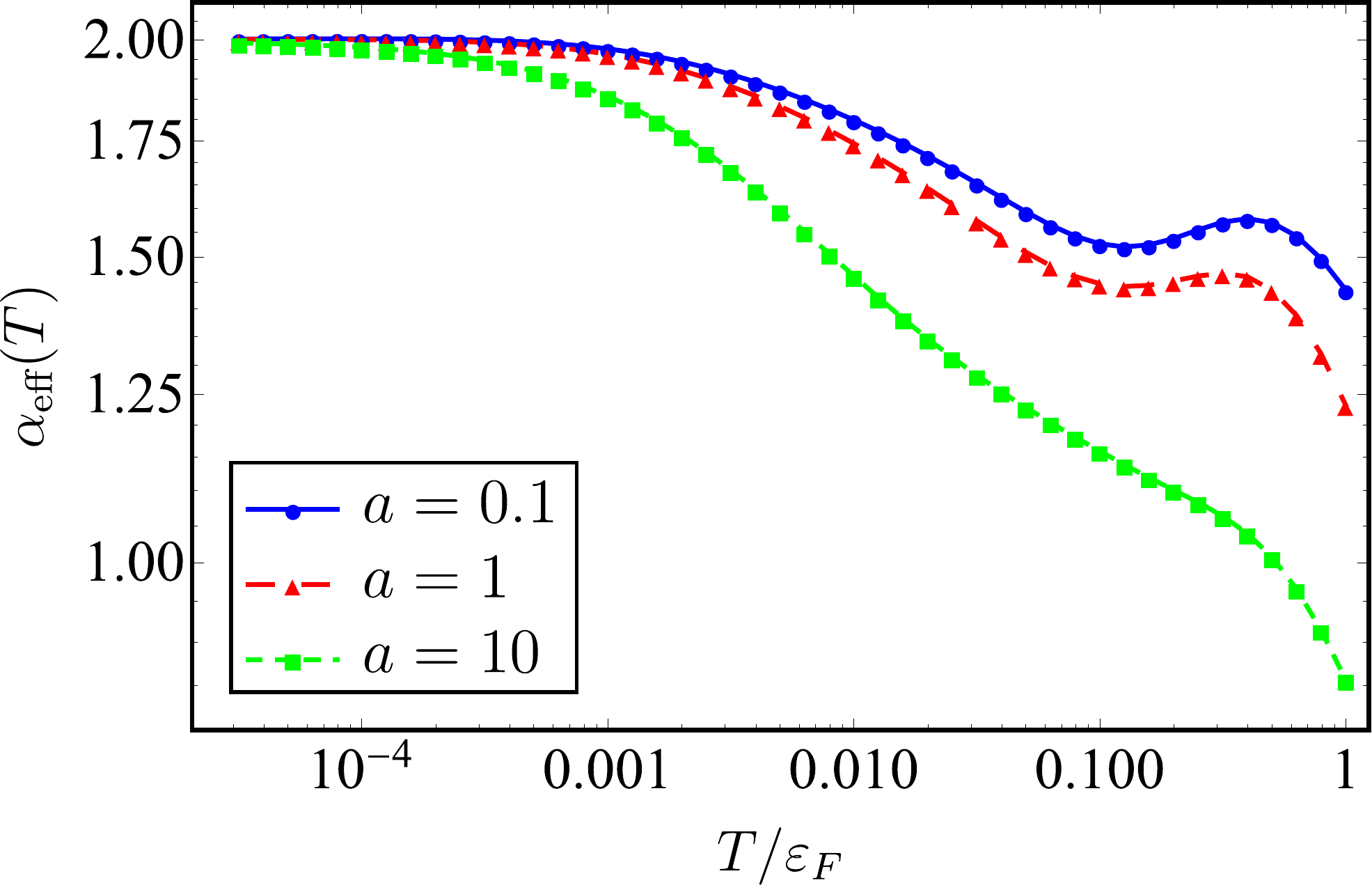}

\caption{(Color online) Dependence of the effective exponent $\alpha_\mathrm{eff}(T)$ of the electrical resistivity $\Delta\rho(T)\sim T^{\alpha_\mathrm{eff}(T)}$
on the nematic mass $r_{0}=r_{0,c}+\delta(T)$ for $\delta(T)=a(T/\varepsilon_F)$. We fix here the impurity scattering and the elastic coupling according to $\kappa_\text{imp}=10^{-1}$ and $\kappa_\text{latt}=10^{-1}$, respectively. In the low-temperature regime, one can notice that the behavior of $\alpha_\mathrm{eff}(T)$ is not altered by the value of the constant $a$.}
\label{Appendix_Resistivity} 
\end{figure}

As discussed in the main text, to reach numerical convergence we
included the temperature dependence of the effective nematic mass at 
the nematic QCP, \textit{i.e.}, we set $r_{0}=r_{0,c}+\delta(T)$. This is a reasonable assumption, since the correlation length only diverges at the QCP. Since a full self-consistent calculation of $\delta(T)$ is beyond the scope of our Boltzmann-equation calculations, here we employ a phenomenological approach. Motivated by the experimental observations \cite{Kuo-S(2016),Bohmer-CRP(2016)} that the nematic susceptibility in iron-based superconductors
displays an approximate Curie-Weiss behavior, $\chi_{\mathrm{nem}}\propto(T-T_{\mathrm{nem}})^{-1}$, we set near the QCP (where $T_{\mathrm{nem}}=0$) $\delta(T)=a(T/\varepsilon_F)$, where $a$ is a dimensionless constant. As shown in 
Fig. \ref{Appendix_Resistivity}, and already anticipated by the analytic results in Eqs. \eqref{Eq_Exact_ResA} 
and \eqref{Eq_Exact_ResB}, the value of $a$ does not change the low-temperature behavior of $\Delta\rho(T)$. It does affect how the effective exponent $\alpha_\mathrm{eff}(T)$ behaves at high temperatures, which is not unexpected, since at high temperatures the resistivity exponent is not universal.


%

\end{document}